% This version 28-Mar-2014
% Last edited by MD

\documentclass[a4paper, traditabstract, longauth]{aa} 
%\documentclass[referee, traditabstract, longauth]{aa} 

% for the abstract without structuration 
% (traditional abstract) 
%
\usepackage{graphicx}
\usepackage[usenames]{color}

\usepackage[breaklinks, colorlinks, citecolor=blue]{hyperref}

\def\all2103resultspapers{\nocite{planck2013-p01, planck2013-p02, planck2013-p02a, planck2013-p02d, planck2013-p02b, planck2013-p03, planck2013-p03c, planck2013-p03f, planck2013-p03d, planck2013-p03e, planck2013-p01a, planck2013-p06, planck2013-p03a, planck2013-pip88, planck2013-p08, planck2013-p11, planck2013-p12, planck2013-p13, planck2013-p14, planck2013-p15, planck2013-p05b, planck2013-p17, planck2013-p09, planck2013-p09a, planck2013-p20, planck2013-p19, planck2013-pipaberration, planck2013-p05, planck2013-p05a, planck2013-pip56, planck2013-p06b}}

\usepackage[nonamebreak]{natbib}
\usepackage{longtable,lscape}
\usepackage{txfonts}
\usepackage{amsfonts}
\usepackage{amssymb}
\usepackage{newlfont}
\usepackage{textcomp}
\usepackage{times}
\usepackage{units}
\usepackage{paralist}
\usepackage{txfonts}
\usepackage{sidecap}

\usepackage{fixltx2e}

\def\setsymbol#1#2{\expandafter\def\csname #1\endcsname{#2}}
\def\getsymbol#1{\csname #1\endcsname}

%-----------------------------------------------------------------------
% Planck
%-----------------------------------------------------------------------
\def\Planck{\textit{Planck}}

%-----------------------------------------------------------------------
% The Planck Helium-4 JT cooler
%-----------------------------------------------------------------------

%-----------------------------------------------------------------------
% To include all Planck Early Results papers in the reference lists
%-----------------------------------------------------------------------

%-----------------------------------------------------------------------
% To include all Planck 2013 Results papers in the reference lists
%-----------------------------------------------------------------------
\def\all2013resultspapers{\nocite{planck2013-p01, planck2013-p02, planck2013-p02a, planck2013-p02d, planck2013-p02b, planck2013-p03, planck2013-p03c, planck2013-p03f, planck2013-p03d, planck2013-p03e, planck2013-p01a, planck2013-p06, planck2013-p03a, planck2013-pip88, planck2013-p08, planck2013-p11, planck2013-p12, planck2013-p13, planck2013-p14, planck2013-p15, planck2013-p05b, planck2013-p17, planck2013-p09, planck2013-p09a, planck2013-p20, planck2013-p19, planck2013-pipaberration, planck2013-p05, planck2013-p05a, planck2013-pip56, planck2013-p06b}}

%-----------------------------------------------------------------------
% Tables
%-----------------------------------------------------------------------
\newbox\tablebox    \newdimen\tablewidth
\def\leaderfil{\leaders\hbox to 5pt{\hss.\hss}\hfil}
%
% use the following definition of \endPlancktable for ApJ style notes to tables, set to the 
%         width of the table
% \def\endPlancktable{\tablewidth=\wd\tablebox 
%
% use the following definitions of \endPlancktable and \endPlancktablewide for A&A style notes 
% set to one-column  or full-page width, respectively
\def\endPlancktable{\tablewidth=\columnwidth 
    $$\hss\copy\tablebox\hss$$
    \vskip-\lastskip\vskip -2pt}
\def\endPlancktablewide{\tablewidth=\textwidth 
    $$\hss\copy\tablebox\hss$$
    \vskip-\lastskip\vskip -2pt}
\def\tablenote#1 #2\par{\begingroup \parindent=0.8em
    \abovedisplayshortskip=0pt\belowdisplayshortskip=0pt
    \noindent
    $$\hss\vbox{\hsize\tablewidth \hangindent=\parindent \hangafter=1 \noindent
    \hbox to \parindent{$^#1$\hss}\strut#2\strut\par}\hss$$
    \endgroup}
\def\doubleline{\vskip 3pt\hrule \vskip 1.5pt \hrule \vskip 5pt}

%-----------------------------------------------------------------------
% useful macros
%-----------------------------------------------------------------------
%
\def\L2{\ifmmode L_2\else $L_2$\fi}

\def\DeltaT{\ifmmode \Delta T\else $\Delta T$\fi}
\def\deltat{\ifmmode \Delta t\else $\Delta t$\fi}
\def\fknee{\ifmmode f_{\rm knee}\else $f_{\rm knee}$\fi}
\def\Fmax{\ifmmode F_{\rm max}\else $F_{\rm max}$\fi}
\def\solar{\ifmmode{\rm M}_{\mathord\odot}\else${\rm M}_{\mathord\odot}$\fi}
\def\Msolar{\ifmmode{\rm M}_{\mathord\odot}\else${\rm M}_{\mathord\odot}$\fi}
\def\Lsolar{\ifmmode{\rm L}_{\mathord\odot}\else${\rm L}_{\mathord\odot}$\fi}

\def\inv{\ifmmode^{-1}\else$^{-1}$\fi}
\def\mo{\ifmmode^{-1}\else$^{-1}$\fi}
\def\sup#1{\ifmmode ^{\rm #1}\else $^{\rm #1}$\fi}
\def\expo#1{\ifmmode \times 10^{#1}\else $\times 10^{#1}$\fi}
\def\,{\thinspace}
\def\lsim{\mathrel{\raise .4ex\hbox{\rlap{$<$}\lower 1.2ex\hbox{$\sim$}}}}
\def\gsim{\mathrel{\raise .4ex\hbox{\rlap{$>$}\lower 1.2ex\hbox{$\sim$}}}}

\def\simprop{\mathrel{\raise .4ex\hbox{\rlap{$\propto$}\lower 1.2ex\hbox{$\sim$}}}}
\def\deg{\ifmmode^\circ\else$^\circ$\fi}
\def\pdeg{\ifmmode $\setbox0=\hbox{$^{\circ}$}\rlap{\hskip.11\wd0 .}$^{\circ}
          \else \setbox0=\hbox{$^{\circ}$}\rlap{\hskip.11\wd0 .}$^{\circ}$\fi}
\def\arcs{\ifmmode {^{\scriptstyle\prime\prime}}
          \else $^{\scriptstyle\prime\prime}$\fi}
\def\arcm{\ifmmode {^{\scriptstyle\prime}}
          \else $^{\scriptstyle\prime}$\fi}
\newdimen\sa  \newdimen\sb
\def\parcs{\sa=.07em \sb=.03em
     \ifmmode \hbox{\rlap{.}}^{\scriptstyle\prime\kern -\sb\prime}\hbox{\kern -\sa}
     \else \rlap{.}$^{\scriptstyle\prime\kern -\sb\prime}$\kern -\sa\fi}
\def\parcm{\sa=.08em \sb=.03em
     \ifmmode \hbox{\rlap{.}\kern\sa}^{\scriptstyle\prime}\hbox{\kern-\sb}
     \else \rlap{.}\kern\sa$^{\scriptstyle\prime}$\kern-\sb\fi}
\def\ra[#1 #2 #3.#4]{#1\sup{h}#2\sup{m}#3\sup{s}\llap.#4}
\def\dec[#1 #2 #3.#4]{#1\deg#2\arcm#3\arcs\llap.#4}
\def\deco[#1 #2 #3]{#1\deg#2\arcm#3\arcs}
\def\rra[#1 #2]{#1\sup{h}#2\sup{m}}

\def\dots{\relax\ifmmode \ldots\else $\ldots$\fi}
%
%-----------------------------------------------------------------------
% units
%-----------------------------------------------------------------------
%
\def\WHzsr{\ifmmode $W\,Hz\mo\,sr\mo$\else W\,Hz\mo\,sr\mo\fi}
\def\mHz{\ifmmode $\,mHz$\else \,mHz\fi}
\def\GHz{\ifmmode $\,GHz$\else \,GHz\fi}
\def\mKs{\ifmmode $\,mK\,s$^{1/2}\else \,mK\,s$^{1/2}$\fi}
\def\muKs{\ifmmode \,\mu$K\,s$^{1/2}\else \,$\mu$K\,s$^{1/2}$\fi}
\def\muKRJs{\ifmmode \,\mu$K$_{\rm RJ}$\,s$^{1/2}\else \,$\mu$K$_{\rm RJ}$\,s$^{1/2}$\fi}
\def\muKHz{\ifmmode \,\mu$K\,Hz$^{-1/2}\else \,$\mu$K\,Hz$^{-1/2}$\fi}
\def\MJysr{\ifmmode \,$MJy\,sr\mo$\else \,MJy\,sr\mo\fi}
\def\MJysrmK{\ifmmode \,$MJy\,sr\mo$\,mK$_{\rm CMB}\mo\else \,MJy\,sr\mo\,mK$_{\rm CMB}\mo$\fi}
\def\microns{\ifmmode \,\mu$m$\else \,$\mu$m\fi}

\def\muK{\ifmmode \,\mu$K$\else \,$\mu$\hbox{K}\fi}
\def\microK{\ifmmode \,\mu$K$\else \,$\mu$\hbox{K}\fi}
\def\muW{\ifmmode \,\mu$W$\else \,$\mu$\hbox{W}\fi}
\def\kms{\ifmmode $\,km\,s$^{-1}\else \,km\,s$^{-1}$\fi}
\def\kmsMpc{\ifmmode $\,\kms\,Mpc\mo$\else \,\kms\,Mpc\mo\fi}
%
%
%----------------------------------------------------------------------

% LFI Center Frequency

\setsymbol{LFI:center:frequency:70GHz:units}{70.3\,GHz}
\setsymbol{LFI:center:frequency:44GHz:units}{44.1\,GHz}
\setsymbol{LFI:center:frequency:30GHz:units}{28.5\,GHz}

\setsymbol{LFI:center:frequency:70GHz}{70.3}
\setsymbol{LFI:center:frequency:44GHz}{44.1}
\setsymbol{LFI:center:frequency:30GHz}{28.5}

\setsymbol{LFI:center:frequency:LFI18:Rad:M:units}{71.7\GHz}
\setsymbol{LFI:center:frequency:LFI19:Rad:M:units}{67.5\GHz}
\setsymbol{LFI:center:frequency:LFI20:Rad:M:units}{69.2\GHz}
\setsymbol{LFI:center:frequency:LFI21:Rad:M:units}{70.4\GHz}
\setsymbol{LFI:center:frequency:LFI22:Rad:M:units}{71.5\GHz}
\setsymbol{LFI:center:frequency:LFI23:Rad:M:units}{70.8\GHz}
\setsymbol{LFI:center:frequency:LFI24:Rad:M:units}{44.4\GHz}
\setsymbol{LFI:center:frequency:LFI25:Rad:M:units}{44.0\GHz}
\setsymbol{LFI:center:frequency:LFI26:Rad:M:units}{43.9\GHz}
\setsymbol{LFI:center:frequency:LFI27:Rad:M:units}{28.3\GHz}
\setsymbol{LFI:center:frequency:LFI28:Rad:M:units}{28.8\GHz}
\setsymbol{LFI:center:frequency:LFI18:Rad:S:units}{70.1\GHz}
\setsymbol{LFI:center:frequency:LFI19:Rad:S:units}{69.6\GHz}
\setsymbol{LFI:center:frequency:LFI20:Rad:S:units}{69.5\GHz}
\setsymbol{LFI:center:frequency:LFI21:Rad:S:units}{69.5\GHz}
\setsymbol{LFI:center:frequency:LFI22:Rad:S:units}{72.8\GHz}
\setsymbol{LFI:center:frequency:LFI23:Rad:S:units}{71.3\GHz}
\setsymbol{LFI:center:frequency:LFI24:Rad:S:units}{44.1\GHz}
\setsymbol{LFI:center:frequency:LFI25:Rad:S:units}{44.1\GHz}
\setsymbol{LFI:center:frequency:LFI26:Rad:S:units}{44.1\GHz}
\setsymbol{LFI:center:frequency:LFI27:Rad:S:units}{28.5\GHz}
\setsymbol{LFI:center:frequency:LFI28:Rad:S:units}{28.2\GHz}

\setsymbol{LFI:center:frequency:LFI18:Rad:M}{71.7}
\setsymbol{LFI:center:frequency:LFI19:Rad:M}{67.5}
\setsymbol{LFI:center:frequency:LFI20:Rad:M}{69.2}
\setsymbol{LFI:center:frequency:LFI21:Rad:M}{70.4}
\setsymbol{LFI:center:frequency:LFI22:Rad:M}{71.5}
\setsymbol{LFI:center:frequency:LFI23:Rad:M}{70.8}
\setsymbol{LFI:center:frequency:LFI24:Rad:M}{44.4}
\setsymbol{LFI:center:frequency:LFI25:Rad:M}{44.0}
\setsymbol{LFI:center:frequency:LFI26:Rad:M}{43.9}
\setsymbol{LFI:center:frequency:LFI27:Rad:M}{28.3}
\setsymbol{LFI:center:frequency:LFI28:Rad:M}{28.8}
\setsymbol{LFI:center:frequency:LFI18:Rad:S}{70.1}
\setsymbol{LFI:center:frequency:LFI19:Rad:S}{69.6}
\setsymbol{LFI:center:frequency:LFI20:Rad:S}{69.5}
\setsymbol{LFI:center:frequency:LFI21:Rad:S}{69.5}
\setsymbol{LFI:center:frequency:LFI22:Rad:S}{72.8}
\setsymbol{LFI:center:frequency:LFI23:Rad:S}{71.3}
\setsymbol{LFI:center:frequency:LFI24:Rad:S}{44.1}
\setsymbol{LFI:center:frequency:LFI25:Rad:S}{44.1}
\setsymbol{LFI:center:frequency:LFI26:Rad:S}{44.1}
\setsymbol{LFI:center:frequency:LFI27:Rad:S}{28.5}
\setsymbol{LFI:center:frequency:LFI28:Rad:S}{28.2}

% LFI White Noise Sensitivity, \delta T_{\rm RJ}

\setsymbol{LFI:white:noise:sensitivity:70GHz:units}{134.7\muKs}
\setsymbol{LFI:white:noise:sensitivity:44GHz:units}{164.7\muKs}
\setsymbol{LFI:white:noise:sensitivity:30GHz:units}{143.4\muKs}

\setsymbol{LFI:white:noise:sensitivity:70GHz}{134.7}
\setsymbol{LFI:white:noise:sensitivity:44GHz}{164.7}
\setsymbol{LFI:white:noise:sensitivity:30GHz}{143.4}

%***************the following are in \delta T_{\rm CMB} units!****************
%***************this segment should be revised********************************

\setsymbol{LFI:white:noise:sensitivity:LFI18:Rad:M:units}{512.0\muKs}
\setsymbol{LFI:white:noise:sensitivity:LFI19:Rad:M:units}{581.4\muKs}
\setsymbol{LFI:white:noise:sensitivity:LFI20:Rad:M:units}{590.8\muKs}
\setsymbol{LFI:white:noise:sensitivity:LFI21:Rad:M:units}{455.2\muKs}
\setsymbol{LFI:white:noise:sensitivity:LFI22:Rad:M:units}{492.0\muKs}
\setsymbol{LFI:white:noise:sensitivity:LFI23:Rad:M:units}{507.7\muKs}
\setsymbol{LFI:white:noise:sensitivity:LFI24:Rad:M:units}{462.2\muKs}
\setsymbol{LFI:white:noise:sensitivity:LFI25:Rad:M:units}{413.6\muKs}
\setsymbol{LFI:white:noise:sensitivity:LFI26:Rad:M:units}{478.6\muKs}
\setsymbol{LFI:white:noise:sensitivity:LFI27:Rad:M:units}{277.7\muKs}
\setsymbol{LFI:white:noise:sensitivity:LFI28:Rad:M:units}{312.3\muKs}
\setsymbol{LFI:white:noise:sensitivity:LFI18:Rad:S:units}{465.7\muKs}
\setsymbol{LFI:white:noise:sensitivity:LFI19:Rad:S:units}{555.6\muKs}
\setsymbol{LFI:white:noise:sensitivity:LFI20:Rad:S:units}{623.2\muKs}
\setsymbol{LFI:white:noise:sensitivity:LFI21:Rad:S:units}{564.1\muKs}
\setsymbol{LFI:white:noise:sensitivity:LFI22:Rad:S:units}{534.4\muKs}
\setsymbol{LFI:white:noise:sensitivity:LFI23:Rad:S:units}{542.4\muKs}
\setsymbol{LFI:white:noise:sensitivity:LFI24:Rad:S:units}{399.2\muKs}
\setsymbol{LFI:white:noise:sensitivity:LFI25:Rad:S:units}{392.6\muKs}
\setsymbol{LFI:white:noise:sensitivity:LFI26:Rad:S:units}{418.6\muKs}
\setsymbol{LFI:white:noise:sensitivity:LFI27:Rad:S:units}{302.9\muKs}
\setsymbol{LFI:white:noise:sensitivity:LFI28:Rad:S:units}{285.3\muKs}

\setsymbol{LFI:white:noise:sensitivity:LFI18:Rad:M}{512.0}
\setsymbol{LFI:white:noise:sensitivity:LFI19:Rad:M}{581.4}
\setsymbol{LFI:white:noise:sensitivity:LFI20:Rad:M}{590.8}
\setsymbol{LFI:white:noise:sensitivity:LFI21:Rad:M}{455.2}
\setsymbol{LFI:white:noise:sensitivity:LFI22:Rad:M}{492.0}
\setsymbol{LFI:white:noise:sensitivity:LFI23:Rad:M}{507.7}
\setsymbol{LFI:white:noise:sensitivity:LFI24:Rad:M}{462.2}
\setsymbol{LFI:white:noise:sensitivity:LFI25:Rad:M}{413.6}
\setsymbol{LFI:white:noise:sensitivity:LFI26:Rad:M}{478.6}
\setsymbol{LFI:white:noise:sensitivity:LFI27:Rad:M}{277.7}
\setsymbol{LFI:white:noise:sensitivity:LFI28:Rad:M}{312.3}
\setsymbol{LFI:white:noise:sensitivity:LFI18:Rad:S}{465.7}
\setsymbol{LFI:white:noise:sensitivity:LFI19:Rad:S}{555.6}
\setsymbol{LFI:white:noise:sensitivity:LFI20:Rad:S}{623.2}
\setsymbol{LFI:white:noise:sensitivity:LFI21:Rad:S}{564.1}
\setsymbol{LFI:white:noise:sensitivity:LFI22:Rad:S}{534.4}
\setsymbol{LFI:white:noise:sensitivity:LFI23:Rad:S}{542.4}
\setsymbol{LFI:white:noise:sensitivity:LFI24:Rad:S}{399.2}
\setsymbol{LFI:white:noise:sensitivity:LFI25:Rad:S}{392.6}
\setsymbol{LFI:white:noise:sensitivity:LFI26:Rad:S}{418.6}
\setsymbol{LFI:white:noise:sensitivity:LFI27:Rad:S}{302.9}
\setsymbol{LFI:white:noise:sensitivity:LFI28:Rad:S}{285.3}

% LFI Knee Frequency

\setsymbol{LFI:knee:frequency:70GHz:units}{29.5\mHz}
\setsymbol{LFI:knee:frequency:44GHz:units}{56.2\mHz}
\setsymbol{LFI:knee:frequency:30GHz:units}{113.7\mHz}

\setsymbol{LFI:knee:frequency:70GHz}{29.5}
\setsymbol{LFI:knee:frequency:44GHz}{56.2}
\setsymbol{LFI:knee:frequency:30GHz}{113.7}

\setsymbol{LFI:knee:frequency:LFI18:Rad:M:units}{16.3\mHz}
\setsymbol{LFI:knee:frequency:LFI19:Rad:M:units}{15.1\mHz}
\setsymbol{LFI:knee:frequency:LFI20:Rad:M:units}{18.7\mHz}
\setsymbol{LFI:knee:frequency:LFI21:Rad:M:units}{37.2\mHz}
\setsymbol{LFI:knee:frequency:LFI22:Rad:M:units}{12.7\mHz}
\setsymbol{LFI:knee:frequency:LFI23:Rad:M:units}{34.6\mHz}
\setsymbol{LFI:knee:frequency:LFI24:Rad:M:units}{46.2\mHz}
\setsymbol{LFI:knee:frequency:LFI25:Rad:M:units}{24.9\mHz}
\setsymbol{LFI:knee:frequency:LFI26:Rad:M:units}{67.6\mHz}
\setsymbol{LFI:knee:frequency:LFI27:Rad:M:units}{187.4\mHz}
\setsymbol{LFI:knee:frequency:LFI28:Rad:M:units}{122.2\mHz}
\setsymbol{LFI:knee:frequency:LFI18:Rad:S:units}{17.7\mHz}
\setsymbol{LFI:knee:frequency:LFI19:Rad:S:units}{22.0\mHz}
\setsymbol{LFI:knee:frequency:LFI20:Rad:S:units}{8.7\mHz}
\setsymbol{LFI:knee:frequency:LFI21:Rad:S:units}{25.9\mHz}
\setsymbol{LFI:knee:frequency:LFI22:Rad:S:units}{15.8\mHz}
\setsymbol{LFI:knee:frequency:LFI23:Rad:S:units}{129.8\mHz}
\setsymbol{LFI:knee:frequency:LFI24:Rad:S:units}{100.9\mHz}
\setsymbol{LFI:knee:frequency:LFI25:Rad:S:units}{38.9\mHz}
\setsymbol{LFI:knee:frequency:LFI26:Rad:S:units}{58.9\mHz}
\setsymbol{LFI:knee:frequency:LFI27:Rad:S:units}{104.4\mHz}
\setsymbol{LFI:knee:frequency:LFI28:Rad:S:units}{40.7\mHz}

\setsymbol{LFI:knee:frequency:LFI18:Rad:M}{16.3}
\setsymbol{LFI:knee:frequency:LFI19:Rad:M}{15.1}
\setsymbol{LFI:knee:frequency:LFI20:Rad:M}{18.7}
\setsymbol{LFI:knee:frequency:LFI21:Rad:M}{37.2}
\setsymbol{LFI:knee:frequency:LFI22:Rad:M}{12.7}
\setsymbol{LFI:knee:frequency:LFI23:Rad:M}{34.6}
\setsymbol{LFI:knee:frequency:LFI24:Rad:M}{46.2}
\setsymbol{LFI:knee:frequency:LFI25:Rad:M}{24.9}
\setsymbol{LFI:knee:frequency:LFI26:Rad:M}{67.6}
\setsymbol{LFI:knee:frequency:LFI27:Rad:M}{187.4}
\setsymbol{LFI:knee:frequency:LFI28:Rad:M}{122.2}
\setsymbol{LFI:knee:frequency:LFI18:Rad:S}{17.7}
\setsymbol{LFI:knee:frequency:LFI19:Rad:S}{22.0}
\setsymbol{LFI:knee:frequency:LFI20:Rad:S}{8.7}
\setsymbol{LFI:knee:frequency:LFI21:Rad:S}{25.9}
\setsymbol{LFI:knee:frequency:LFI22:Rad:S}{15.8}
\setsymbol{LFI:knee:frequency:LFI23:Rad:S}{129.8}
\setsymbol{LFI:knee:frequency:LFI24:Rad:S}{100.9}
\setsymbol{LFI:knee:frequency:LFI25:Rad:S}{38.9}
\setsymbol{LFI:knee:frequency:LFI26:Rad:S}{58.9}
\setsymbol{LFI:knee:frequency:LFI27:Rad:S}{104.4}
\setsymbol{LFI:knee:frequency:LFI28:Rad:S}{40.7}

% LFI low frequency noise slope

\setsymbol{LFI:slope:70GHz:units}{$-1.03$\mHz}
\setsymbol{LFI:slope:44GHz:units}{$-0.89$\mHz}
\setsymbol{LFI:slope:30GHz:units}{$-0.87$\mHz}

\setsymbol{LFI:slope:70GHz}{$-1.03$}
\setsymbol{LFI:slope:44GHz}{$-0.89$}
\setsymbol{LFI:slope:30GHz}{$-0.87$}

\setsymbol{LFI:slope:LFI18:Rad:M:units}{$-1.04$\mHz}
\setsymbol{LFI:slope:LFI19:Rad:M:units}{$-1.09$\mHz}
\setsymbol{LFI:slope:LFI20:Rad:M:units}{$-0.69$\mHz}
\setsymbol{LFI:slope:LFI21:Rad:M:units}{$-1.56$\mHz}
\setsymbol{LFI:slope:LFI22:Rad:M:units}{$-1.01$\mHz}
\setsymbol{LFI:slope:LFI23:Rad:M:units}{$-0.96$\mHz}
\setsymbol{LFI:slope:LFI24:Rad:M:units}{$-0.83$\mHz}
\setsymbol{LFI:slope:LFI25:Rad:M:units}{$-0.91$\mHz}
\setsymbol{LFI:slope:LFI26:Rad:M:units}{$-0.95$\mHz}
\setsymbol{LFI:slope:LFI27:Rad:M:units}{$-0.87$\mHz}
\setsymbol{LFI:slope:LFI28:Rad:M:units}{$-0.88$\mHz}
\setsymbol{LFI:slope:LFI18:Rad:S:units}{$-1.15$\mHz}
\setsymbol{LFI:slope:LFI19:Rad:S:units}{$-1.00$\mHz}
\setsymbol{LFI:slope:LFI20:Rad:S:units}{$-0.95$\mHz}
\setsymbol{LFI:slope:LFI21:Rad:S:units}{$-0.92$\mHz}
\setsymbol{LFI:slope:LFI22:Rad:S:units}{$-1.01$\mHz}
\setsymbol{LFI:slope:LFI23:Rad:S:units}{$-0.95$\mHz}
\setsymbol{LFI:slope:LFI24:Rad:S:units}{$-0.73$\mHz}
\setsymbol{LFI:slope:LFI25:Rad:S:units}{$-1.16$\mHz}
\setsymbol{LFI:slope:LFI26:Rad:S:units}{$-0.79$\mHz}
\setsymbol{LFI:slope:LFI27:Rad:S:units}{$-0.82$\mHz}
\setsymbol{LFI:slope:LFI28:Rad:S:units}{$-0.91$\mHz}

\setsymbol{LFI:slope:LFI18:Rad:M}{$-1.04$}
\setsymbol{LFI:slope:LFI19:Rad:M}{$-1.09$}
\setsymbol{LFI:slope:LFI20:Rad:M}{$-0.69$}
\setsymbol{LFI:slope:LFI21:Rad:M}{$-1.56$}
\setsymbol{LFI:slope:LFI22:Rad:M}{$-1.01$}
\setsymbol{LFI:slope:LFI23:Rad:M}{$-0.96$}
\setsymbol{LFI:slope:LFI24:Rad:M}{$-0.83$}
\setsymbol{LFI:slope:LFI25:Rad:M}{$-0.91$}
\setsymbol{LFI:slope:LFI26:Rad:M}{$-0.95$}
\setsymbol{LFI:slope:LFI27:Rad:M}{$-0.87$}
\setsymbol{LFI:slope:LFI28:Rad:M}{$-0.88$}
\setsymbol{LFI:slope:LFI18:Rad:S}{$-1.15$}
\setsymbol{LFI:slope:LFI19:Rad:S}{$-1.00$}
\setsymbol{LFI:slope:LFI20:Rad:S}{$-0.95$}
\setsymbol{LFI:slope:LFI21:Rad:S}{$-0.92$}
\setsymbol{LFI:slope:LFI22:Rad:S}{$-1.01$}
\setsymbol{LFI:slope:LFI23:Rad:S}{$-0.95$}
\setsymbol{LFI:slope:LFI24:Rad:S}{$-0.73$}
\setsymbol{LFI:slope:LFI25:Rad:S}{$-1.16$}
\setsymbol{LFI:slope:LFI26:Rad:S}{$-0.79$}
\setsymbol{LFI:slope:LFI27:Rad:S}{$-0.82$}
\setsymbol{LFI:slope:LFI28:Rad:S}{$-0.91$}

% LFI Beam FWHM

\setsymbol{LFI:FWHM:70GHz:units}{13\parcm01}
\setsymbol{LFI:FWHM:44GHz:units}{27\parcm92}
\setsymbol{LFI:FWHM:30GHz:units}{32\parcm65}

\setsymbol{LFI:FWHM:70GHz}{13.01}
\setsymbol{LFI:FWHM:44GHz}{27.92}
\setsymbol{LFI:FWHM:30GHz}{32.65}

\setsymbol{LFI:FWHM:LFI18:units}{13\parcm39}
\setsymbol{LFI:FWHM:LFI19:units}{13\parcm01}
\setsymbol{LFI:FWHM:LFI20:units}{12\parcm75}
\setsymbol{LFI:FWHM:LFI21:units}{12\parcm74}
\setsymbol{LFI:FWHM:LFI22:units}{12\parcm87}
\setsymbol{LFI:FWHM:LFI23:units}{13\parcm27}
\setsymbol{LFI:FWHM:LFI24:units}{22\parcm98}
\setsymbol{LFI:FWHM:LFI25:units}{30\parcm46}
\setsymbol{LFI:FWHM:LFI26:units}{30\parcm31}
\setsymbol{LFI:FWHM:LFI27:units}{32\parcm65}
\setsymbol{LFI:FWHM:LFI28:units}{32\parcm66}

\setsymbol{LFI:FWHM:LFI18}{13.39}
\setsymbol{LFI:FWHM:LFI19}{13.01}
\setsymbol{LFI:FWHM:LFI20}{12.75}
\setsymbol{LFI:FWHM:LFI21}{12.74}
\setsymbol{LFI:FWHM:LFI22}{12.87}
\setsymbol{LFI:FWHM:LFI23}{13.27}
\setsymbol{LFI:FWHM:LFI24}{22.98}
\setsymbol{LFI:FWHM:LFI25}{30.46}
\setsymbol{LFI:FWHM:LFI26}{30.31}
\setsymbol{LFI:FWHM:LFI27}{32.65}
\setsymbol{LFI:FWHM:LFI28}{32.66}

% LFI Beam FWHM Uncertainty
% When uncertainties are routinely available for all quantities, we'll likely change the format to build them into 
% the \setsymbol command.  For now, this will be a bit easier.

%\setsymbol{LFI:FWHM:uncertainty:70GHz}{TBD\arcm}
%\setsymbol{LFI:FWHM:uncertainty:44GHz}{TBD\arcm}
%\setsymbol{LFI:FWHM:uncertainty:30GHz}{TBD\arcm}

\setsymbol{LFI:FWHM:uncertainty:LFI18:units}{0.170\arcm}
\setsymbol{LFI:FWHM:uncertainty:LFI19:units}{0.174\arcm}
\setsymbol{LFI:FWHM:uncertainty:LFI20:units}{0.170\arcm}
\setsymbol{LFI:FWHM:uncertainty:LFI21:units}{0.156\arcm}
\setsymbol{LFI:FWHM:uncertainty:LFI22:units}{0.164\arcm}
\setsymbol{LFI:FWHM:uncertainty:LFI23:units}{0.171\arcm}
\setsymbol{LFI:FWHM:uncertainty:LFI24:units}{0.652\arcm}
\setsymbol{LFI:FWHM:uncertainty:LFI25:units}{1.075\arcm}
\setsymbol{LFI:FWHM:uncertainty:LFI26:units}{1.131\arcm}
\setsymbol{LFI:FWHM:uncertainty:LFI27:units}{1.266\arcm}
\setsymbol{LFI:FWHM:uncertainty:LFI28:units}{1.287\arcm}

\setsymbol{LFI:FWHM:uncertainty:LFI18}{0.170}
\setsymbol{LFI:FWHM:uncertainty:LFI19}{0.174}
\setsymbol{LFI:FWHM:uncertainty:LFI20}{0.170}
\setsymbol{LFI:FWHM:uncertainty:LFI21}{0.156}
\setsymbol{LFI:FWHM:uncertainty:LFI22}{0.164}
\setsymbol{LFI:FWHM:uncertainty:LFI23}{0.171}
\setsymbol{LFI:FWHM:uncertainty:LFI24}{0.652}
\setsymbol{LFI:FWHM:uncertainty:LFI25}{1.075}
\setsymbol{LFI:FWHM:uncertainty:LFI26}{1.131}
\setsymbol{LFI:FWHM:uncertainty:LFI27}{1.266}
\setsymbol{LFI:FWHM:uncertainty:LFI28}{1.287}

% HFI Center Frequency

\setsymbol{HFI:center:frequency:100GHz:units}{100\,GHz}
\setsymbol{HFI:center:frequency:143GHz:units}{143\,GHz}
\setsymbol{HFI:center:frequency:217GHz:units}{217\,GHz}
\setsymbol{HFI:center:frequency:353GHz:units}{353\,GHz}
\setsymbol{HFI:center:frequency:545GHz:units}{545\,GHz}
\setsymbol{HFI:center:frequency:857GHz:units}{857\,GHz}

\setsymbol{HFI:center:frequency:100GHz}{100}
\setsymbol{HFI:center:frequency:143GHz}{143}
\setsymbol{HFI:center:frequency:217GHz}{217}
\setsymbol{HFI:center:frequency:353GHz}{353}
\setsymbol{HFI:center:frequency:545GHz}{545}
\setsymbol{HFI:center:frequency:857GHz}{857}

% HFI Number of Detectors

\setsymbol{HFI:Ndetectors:100GHz}{8}
\setsymbol{HFI:Ndetectors:143GHz}{11}
\setsymbol{HFI:Ndetectors:217GHz}{12}
\setsymbol{HFI:Ndetectors:353GHz}{12}
\setsymbol{HFI:Ndetectors:545GHz}{3}
\setsymbol{HFI:Ndetectors:857GHz}{4}

% HFI FWHM on maps

\setsymbol{HFI:FWHM:Maps:100GHz:units}{9\parcm88}
\setsymbol{HFI:FWHM:Maps:143GHz:units}{7\parcm18}
\setsymbol{HFI:FWHM:Maps:217GHz:units}{4\parcm87}
\setsymbol{HFI:FWHM:Maps:353GHz:units}{4\parcm65}
\setsymbol{HFI:FWHM:Maps:545GHz:units}{4\parcm72}
\setsymbol{HFI:FWHM:Maps:857GHz:units}{4\parcm39}
\setsymbol{HFI:FWHM:Maps:100GHz}{9.88}
\setsymbol{HFI:FWHM:Maps:143GHz}{7.18}
\setsymbol{HFI:FWHM:Maps:217GHz}{4.87}
\setsymbol{HFI:FWHM:Maps:353GHz}{4.65}
\setsymbol{HFI:FWHM:Maps:545GHz}{4.72}
\setsymbol{HFI:FWHM:Maps:857GHz}{4.39}

% HFI Beam Ellipticity on maps

\setsymbol{HFI:beam:ellipticity:Maps:100GHz}{1.15}
\setsymbol{HFI:beam:ellipticity:Maps:143GHz}{1.01}
\setsymbol{HFI:beam:ellipticity:Maps:217GHz}{1.06}
\setsymbol{HFI:beam:ellipticity:Maps:353GHz}{1.05}
\setsymbol{HFI:beam:ellipticity:Maps:545GHz}{1.14}
\setsymbol{HFI:beam:ellipticity:Maps:857GHz}{1.19}

% HFI optical Beam FWHM from Mars; time response deconvolved: frequency  average of values in table 4 in HFI instrument paper

\setsymbol{HFI:FWHM:Mars:100GHz:units}{9\parcm37}
\setsymbol{HFI:FWHM:Mars:143GHz:units}{7\parcm04}
\setsymbol{HFI:FWHM:Mars:217GHz:units}{4\parcm68}
\setsymbol{HFI:FWHM:Mars:353GHz:units}{4\parcm43}
\setsymbol{HFI:FWHM:Mars:545GHz:units}{3\parcm80}
\setsymbol{HFI:FWHM:Mars:857GHz:units}{3\parcm67}

\setsymbol{HFI:FWHM:Mars:100GHz}{9.37}
\setsymbol{HFI:FWHM:Mars:143GHz}{7.04}
\setsymbol{HFI:FWHM:Mars:217GHz}{4.68}
\setsymbol{HFI:FWHM:Mars:353GHz}{4.43}
\setsymbol{HFI:FWHM:Mars:545GHz}{3.80}
\setsymbol{HFI:FWHM:Mars:857GHz}{3.67}

% HFI optical Beam Ellipticity from Mars; time response deconvolved: frequency average of values in table 4 in HFI instrument paper

\setsymbol{HFI:beam:ellipticity:Mars:100GHz}{1.18}
\setsymbol{HFI:beam:ellipticity:Mars:143GHz}{1.03}
\setsymbol{HFI:beam:ellipticity:Mars:217GHz}{1.14}
\setsymbol{HFI:beam:ellipticity:Mars:353GHz}{1.09}
\setsymbol{HFI:beam:ellipticity:Mars:545GHz}{1.25}
\setsymbol{HFI:beam:ellipticity:Mars:857GHz}{1.03}

% HFI CMB relative calibration accuracy

\setsymbol{HFI:CMB:relative:calibration:100GHz}{$\lsim 1\%$}
\setsymbol{HFI:CMB:relative:calibration:143GHz}{$\lsim 1\%$}
\setsymbol{HFI:CMB:relative:calibration:217GHz}{$\lsim 1\%$}
\setsymbol{HFI:CMB:relative:calibration:353GHz}{$\lsim 1\%$}
\setsymbol{HFI:CMB:relative:calibration:545GHz}{}
\setsymbol{HFI:CMB:relative:calibration:857GHz}{}

% HFI CMB absolute calibration accuracy

\setsymbol{HFI:CMB:absolute:calibration:100GHz}{$\lsim 2\%$}
\setsymbol{HFI:CMB:absolute:calibration:143GHz}{$\lsim 2\%$}
\setsymbol{HFI:CMB:absolute:calibration:217GHz}{$\lsim 2\%$}
\setsymbol{HFI:CMB:absolute:calibration:353GHz}{$\lsim 2\%$}
\setsymbol{HFI:CMB:absolute:calibration:545GHz}{}
\setsymbol{HFI:CMB:absolute:calibration:857GHz}{}

% HFI FIRAS gain calibration accuracy: statistical

\setsymbol{HFI:FIRAS:gain:calibration:accuracy:statistical:100GHz}{}
\setsymbol{HFI:FIRAS:gain:calibration:accuracy:statistical:143GHz}{}
\setsymbol{HFI:FIRAS:gain:calibration:accuracy:statistical:217GHz}{}
\setsymbol{HFI:FIRAS:gain:calibration:accuracy:statistical:353GHz}{2.5\%}
\setsymbol{HFI:FIRAS:gain:calibration:accuracy:statistical:545GHz}{1\%}
\setsymbol{HFI:FIRAS:gain:calibration:accuracy:statistical:857GHz}{0.5\%}

% HFI FIRAS gain calibration accuracy: systematic

\setsymbol{HFI:FIRAS:gain:calibration:accuracy:systematic:100GHz}{}
\setsymbol{HFI:FIRAS:gain:calibration:accuracy:systematic:143GHz}{}
\setsymbol{HFI:FIRAS:gain:calibration:accuracy:systematic:217GHz}{}
\setsymbol{HFI:FIRAS:gain:calibration:accuracy:systematic:353GHz}{}
\setsymbol{HFI:FIRAS:gain:calibration:accuracy:systematic:545GHz}{7\%}
\setsymbol{HFI:FIRAS:gain:calibration:accuracy:systematic:857GHz}{7\%}

% HFI FIRAS zero point accuracy:

\setsymbol{HFI:FIRAS:zero:point:accuracy:100GHz:units}{0.8\MJysr}
\setsymbol{HFI:FIRAS:zero:point:accuracy:143GHz:units}{}
\setsymbol{HFI:FIRAS:zero:point:accuracy:217GHz:units}{}
\setsymbol{HFI:FIRAS:zero:point:accuracy:353GHz:units}{1.4\MJysr}
\setsymbol{HFI:FIRAS:zero:point:accuracy:545GHz:units}{2.2\MJysr}
\setsymbol{HFI:FIRAS:zero:point:accuracy:857GHz:units}{1.7\MJysr}

\setsymbol{HFI:FIRAS:zero:point:accuracy:100GHz}{0.8}
\setsymbol{HFI:FIRAS:zero:point:accuracy:143GHz}{}
\setsymbol{HFI:FIRAS:zero:point:accuracy:217GHz}{}
\setsymbol{HFI:FIRAS:zero:point:accuracy:353GHz}{1.4}
\setsymbol{HFI:FIRAS:zero:point:accuracy:545GHz}{2.2}
\setsymbol{HFI:FIRAS:zero:point:accuracy:857GHz}{1.7}

% HFI diffuse source sensitivity unit conversion

\setsymbol{HFI:unit:conversion:100GHz:units}{0.2415\MJysrmK}
\setsymbol{HFI:unit:conversion:143GHz:units}{0.3694\MJysrmK}
\setsymbol{HFI:unit:conversion:217GHz:units}{0.4811\MJysrmK}
\setsymbol{HFI:unit:conversion:353GHz:units}{0.2883\MJysrmK}
\setsymbol{HFI:unit:conversion:545GHz:units}{0.05826\MJysrmK}
\setsymbol{HFI:unit:conversion:857GHz:units}{0.002238\MJysrmK}

\setsymbol{HFI:unit:conversion:100GHz}{0.2415}
\setsymbol{HFI:unit:conversion:143GHz}{0.3694}
\setsymbol{HFI:unit:conversion:217GHz}{0.4811}
\setsymbol{HFI:unit:conversion:353GHz}{0.2883}
\setsymbol{HFI:unit:conversion:545GHz}{0.05826}
\setsymbol{HFI:unit:conversion:857GHz}{0.002238}

% HFI Colour Correction for \alpha = -2, for V1.01 of the spectral bands

\setsymbol{HFI:colour:correction:alpha=-2:V1.01:100GHz}{0.9893}
\setsymbol{HFI:colour:correction:alpha=-2:V1.01:143GHz}{0.9759}
\setsymbol{HFI:colour:correction:alpha=-2:V1.01:217GHz}{1.0007}
\setsymbol{HFI:colour:correction:alpha=-2:V1.01:353GHz}{1.0028}
\setsymbol{HFI:colour:correction:alpha=-2:V1.01:545GHz}{1.0019}
\setsymbol{HFI:colour:correction:alpha=-2:V1.01:857GHz}{0.9889}

% HFI Colour Correction for \alpha = 0, for V1.01 of the spectral bands

\setsymbol{HFI:colour:correction:alpha=0:V1.01:100GHz}{1.0008}
\setsymbol{HFI:colour:correction:alpha=0:V1.01:143GHz}{1.0148}
\setsymbol{HFI:colour:correction:alpha=0:V1.01:217GHz}{0.9909}
\setsymbol{HFI:colour:correction:alpha=0:V1.01:353GHz}{0.9888}
\setsymbol{HFI:colour:correction:alpha=0:V1.01:545GHz}{0.9878}
\setsymbol{HFI:colour:correction:alpha=0:V1.01:857GHz}{1.0014}

\providecommand{\sorthelp}[1]{}

\def\reff@jnl#1{{\rm#1\/}}
\def\apj{\reff@jnl{ApJ}}       % Astrophysical Journal
\def\apjs{\reff@jnl{ApJS}}     % Astrophysical Journal, Supplement
\def\aaps{\reff@jnl{A\&AS}}    % Astronomy and Astrophysics, Supplement
\def\mnras{\reff@jnl{MNRAS}}   % Monthly Notices of the RAS
\def\prd{\reff@jnl{Phys.\ Rev.\ D}}    % Physical Review D

 % N_side
   % N_pix
   % N_tau: one sided length of kernel

 % transpose
\newcommand{\beq}{\begin{equation}}
\newcommand{\eeq}{\end{equation}}

\newcommand{\be}{\begin{equation}}
\newcommand{\ee}{\end{equation}}
\newcommand{\bea}{\begin{eq}}
\newcommand{\eea}{\end{equation}}

{\begin{enumerate}\setlength{\itemsep}{0mm}}%
{\end{enumerate}}
{\begin{enumerate}\setlength{\itemsep}{0mm}}%
{\end{enumerate}}

\newcommand{\bc}{\begin{center}}
\newcommand{\ec}{\end{center}}
\newcommand{\bi}{\begin{itemize}}
\newcommand{\ei}{\end{itemize}}
\newcommand{\ben}{\begin{enumerate}}
\newcommand{\een}{\end{enumerate}}

\usepackage{mathrsfs}
\usepackage{mathbbol}

\newfont{\gwpfont}{cmssq8 scaled 1000}
\newcommand{\rexcess}{{\gwpfont REXCESS}}

\def\xmm{{\it XMM-Newton}}
\def\Mv {M_\mathrm{500}}
\def\msol {\mathrm{M}_{\odot}}
\def\YX {Y_\mathrm{X}} 
\def \Rv {R_{500}} 
\def\keV {\mathrm{keV}} 

%%%%%%%%%%%%%%%%%%%%%%%%%%%%%%%%%%%%%%%%

\def\Mgv{M_{\mathrm{ g},500}}

\def\YX {Y_{\mathrm{ X}}}

\def\TX {T_{\mathrm{ X}}}

\def\Mv {M_{\mathrm{ 500}}}
\def \Rv {R_{500}}
\def\keV {\mathrm{ keV}}
\def\Yv {Y_{500}}

\def\MYX {$M$--$Y_{\mathrm{ X}}$}

\def\MYX {$\Mv$--$\YX$}

\def\MyMh {$\Mvy$--$\Mvh$}
\def\MyM {$\Mvy$--$\Mv$}
\def\MhYX {$\Mvh$--$\YX$}
\def\YXM {$\YX$--$\Mv$}

\def\YXMh {$\YX$--$\Mvh$}
\def\YMh {$\YSZ$--$\Mvh$}
\def\YM {$\YSZ$--$\Mv$}
\def\YMy {$\YSZ$--$\Mvy$}

\def\YSZ {Y_{500}}
\def\Mv {M_{\mathrm{ 500}}}
\def\Mvy {\Mv^{\YX}}
\def\Mvh {\Mv^{\mathrm{ HE}}}
\def\Rvy {\Rv^{\YX}}

\def\lesssim{\mathrel{\hbox{\rlap{\hbox{\lower4pt\hbox{$\sim$}}}\hbox{$<$}}}}
\def\gtrsim{\mathrel{\hbox{\rlap{\hbox{\lower4pt\hbox{$\sim$}}}\hbox{$>$}}}}

% satellites
\def \xmm {\hbox{\it XMM-Newton}}
\def \chandra {\hbox{\it Chandra}}

\def \planck {\hbox{\it Planck}}

%%%%%%%%%%%%%%%%%%%%%%%%%%%%%%%%%%%%%%%%

%%%%%%%%%%%%%%%%%%%%%%%%%%%%%%%%%%%%%%%%%%%%%%%%%%%%%%%%%%%%

\begin{document}

%This author list corresponds to \title{Author list for SVN P15\_Cosmological\_Constraints\_from\_Clusters, Proj. Ref. 5\_1: Cosmology from Planck SZ cluster counts}
%Prepared by R. Leonardi (rleonardi@sciops.esa.int), ESAC/ESA
%This version is from Wed Mar 20 09:41:50 2013 CET
%\subtitle{There are 254 co-authors in this list}
\author{\small
Planck Collaboration:
P.~A.~R.~Ade\inst{95}
\and
N.~Aghanim\inst{66}
\and
C.~Armitage-Caplan\inst{101}
\and
M.~Arnaud\inst{79}
\and
M.~Ashdown\inst{76, 7}
\and
F.~Atrio-Barandela\inst{21}
\and
J.~Aumont\inst{66}
\and
C.~Baccigalupi\inst{93}
\and
A.~J.~Banday\inst{105, 11}
\and
R.~B.~Barreiro\inst{73}
% ...
}
\institute{\small
APC, AstroParticule et Cosmologie, Universit\'{e} Paris Diderot, CNRS/IN2P3, CEA/lrfu, Observatoire de Paris, Sorbonne Paris Cit\'{e}, 10, rue Alice Domon et L\'{e}onie Duquet, 75205 Paris Cedex 13, France\\
\and
Aalto University Mets\"{a}hovi Radio Observatory, Mets\"{a}hovintie 114, FIN-02540 Kylm\"{a}l\"{a}, Finland\\
\and
Academy of Sciences of Tatarstan, Bauman Str., 20, Kazan, 420111, Republic of Tatarstan, Russia\\
\and
African Institute for Mathematical Sciences, 6-8 Melrose Road, Muizenberg, Cape Town, South Africa\\
\and
Agenzia Spaziale Italiana Science Data Center, c/o ESRIN, via Galileo Galilei, Frascati, Italy\\
\and
Agenzia Spaziale Italiana, Viale Liegi 26, Roma, Italy\\
\and
Astrophysics Group, Cavendish Laboratory, University of Cambridge, J J Thomson Avenue, Cambridge CB3 0HE, U.K.\\
\and
Astrophysics \& Cosmology Research Unit, School of Mathematics, Statistics \& Computer Science, University of KwaZulu-Natal, Westville Campus, Private Bag X54001, Durban 4000, South Africa\\
\and
Atacama Large Millimeter/submillimeter Array, ALMA Santiago Central Offices, Alonso de Cordova 3107, Vitacura, Casilla 763 0355, Santiago, Chile\\
\and
CITA, University of Toronto, 60 St. George St., Toronto, ON M5S 3H8, Canada\\
\and
CNRS, IRAP, 9 Av. colonel Roche, BP 44346, F-31028 Toulouse cedex 4, France\\
\and
California Institute of Technology, Pasadena, California, U.S.A.\\
\and
Centre for Theoretical Cosmology, DAMTP, University of Cambridge, Wilberforce Road, Cambridge CB3 0WA U.K.\\
\and
Centro de Astrof\'{\i}sica, Universidade do Porto, Rua das Estrelas, 4150-762 Porto, Portugal\\
\and
Centro de Estudios de F\'{i}sica del Cosmos de Arag\'{o}n (CEFCA), Plaza San Juan, 1, planta 2, E-44001, Teruel, Spain\\
\and
Computational Cosmology Center, Lawrence Berkeley National Laboratory, Berkeley, California, U.S.A.\\
\and
Consejo Superior de Investigaciones Cient\'{\i}ficas (CSIC), Madrid, Spain\\
\and
DSM/Irfu/SPP, CEA-Saclay, F-91191 Gif-sur-Yvette Cedex, France\\
\and
DTU Space, National Space Institute, Technical University of Denmark, Elektrovej 327, DK-2800 Kgs. Lyngby, Denmark\\
\and
D\'{e}partement de Physique Th\'{e}orique, Universit\'{e} de Gen\`{e}ve, 24, Quai E. Ansermet,1211 Gen\`{e}ve 4, Switzerland\\
\and
Departamento de F\'{\i}sica Fundamental, Facultad de Ciencias, Universidad de Salamanca, 37008 Salamanca, Spain\\
\and
Departamento de F\'{\i}sica, Universidad de Oviedo, Avda. Calvo Sotelo s/n, Oviedo, Spain\\
\and
Department of Astronomy and Astrophysics, University of Toronto, 50 Saint George Street, Toronto, Ontario, Canada\\
\and
Department of Astronomy and Geodesy, Kazan Federal University,  Kremlevskaya Str., 18, Kazan, 420008, Russia\\
\and
Department of Astrophysics/IMAPP, Radboud University Nijmegen, P.O. Box 9010, 6500 GL Nijmegen, The Netherlands\\
\and
Department of Electrical Engineering and Computer Sciences, University of California, Berkeley, California, U.S.A.\\
\and
Department of Physics \& Astronomy, University of British Columbia, 6224 Agricultural Road, Vancouver, British Columbia, Canada\\
\and
Department of Physics and Astronomy, Dana and David Dornsife College of Letter, Arts and Sciences, University of Southern California, Los Angeles, CA 90089, U.S.A.\\
\and
Department of Physics and Astronomy, University College London, London WC1E 6BT, U.K.\\
\and
Department of Physics and Astronomy, University of Sussex, Brighton BN1 9QH, U.K.\\
\and
Department of Physics, Gustaf H\"{a}llstr\"{o}min katu 2a, University of Helsinki, Helsinki, Finland\\
\and
Department of Physics, Princeton University, Princeton, New Jersey, U.S.A.\\
\and
Department of Physics, University of California, Berkeley, California, U.S.A.\\
\and
Department of Physics, University of California, One Shields Avenue, Davis, California, U.S.A.\\
\and
Department of Physics, University of California, Santa Barbara, California, U.S.A.\\
\and
Department of Physics, University of Illinois at Urbana-Champaign, 1110 West Green Street, Urbana, Illinois, U.S.A.\\
\and
Dipartimento di Fisica e Astronomia G. Galilei, Universit\`{a} degli Studi di Padova, via Marzolo 8, 35131 Padova, Italy\\
\and
Dipartimento di Fisica e Scienze della Terra, Universit\`{a} di Ferrara, Via Saragat 1, 44122 Ferrara, Italy\\
\and
Dipartimento di Fisica, Universit\`{a} La Sapienza, P. le A. Moro 2, Roma, Italy\\
\and
Dipartimento di Fisica, Universit\`{a} degli Studi di Milano, Via Celoria, 16, Milano, Italy\\
\and
Dipartimento di Fisica, Universit\`{a} degli Studi di Trieste, via A. Valerio 2, Trieste, Italy\\
\and
Dipartimento di Fisica, Universit\`{a} di Roma Tor Vergata, Via della Ricerca Scientifica, 1, Roma, Italy\\
\and
Discovery Center, Niels Bohr Institute, Blegdamsvej 17, Copenhagen, Denmark\\
\and
Dpto. Astrof\'{i}sica, Universidad de La Laguna (ULL), E-38206 La Laguna, Tenerife, Spain\\
\and
European Southern Observatory, ESO Vitacura, Alonso de Cordova 3107, Vitacura, Casilla 19001, Santiago, Chile\\
\and
European Space Agency, ESAC, Planck Science Office, Camino bajo del Castillo, s/n, Urbanizaci\'{o}n Villafranca del Castillo, Villanueva de la Ca\~{n}ada, Madrid, Spain\\
\and
European Space Agency, ESTEC, Keplerlaan 1, 2201 AZ Noordwijk, The Netherlands\\
\and
Finnish Centre for Astronomy with ESO (FINCA), University of Turku, V\"{a}is\"{a}l\"{a}ntie 20, FIN-21500, Piikki\"{o}, Finland\\
\and
Haverford College Astronomy Department, 370 Lancaster Avenue, Haverford, Pennsylvania, U.S.A.\\
\and
Helsinki Institute of Physics, Gustaf H\"{a}llstr\"{o}min katu 2, University of Helsinki, Helsinki, Finland\\
\and
INAF - Osservatorio Astrofisico di Catania, Via S. Sofia 78, Catania, Italy\\
\and
INAF - Osservatorio Astronomico di Padova, Vicolo dell'Osservatorio 5, Padova, Italy\\
\and
INAF - Osservatorio Astronomico di Roma, via di Frascati 33, Monte Porzio Catone, Italy\\
\and
INAF - Osservatorio Astronomico di Trieste, Via G.B. Tiepolo 11, Trieste, Italy\\
\and
INAF/IASF Bologna, Via Gobetti 101, Bologna, Italy\\
\and
INAF/IASF Milano, Via E. Bassini 15, Milano, Italy\\
\and
INFN, Sezione di Bologna, Via Irnerio 46, I-40126, Bologna, Italy\\
\and
INFN, Sezione di Roma 1, Universit\`{a} di Roma Sapienza, Piazzale Aldo Moro 2, 00185, Roma, Italy\\
\and
IPAG: Institut de Plan\'{e}tologie et d'Astrophysique de Grenoble, Universit\'{e} Joseph Fourier, Grenoble 1 / CNRS-INSU, UMR 5274, Grenoble, F-38041, France\\
\and
ISDC Data Centre for Astrophysics, University of Geneva, ch. d'Ecogia 16, Versoix, Switzerland\\
\and
IUCAA, Post Bag 4, Ganeshkhind, Pune University Campus, Pune 411 007, India\\
\and
Imperial College London, Astrophysics group, Blackett Laboratory, Prince Consort Road, London, SW7 2AZ, U.K.\\
\and
Infrared Processing and Analysis Center, California Institute of Technology, Pasadena, CA 91125, U.S.A.\\
\and
Institut N\'{e}el, CNRS, Universit\'{e} Joseph Fourier Grenoble I, 25 rue des Martyrs, Grenoble, France\\
\and
Institut Universitaire de France, 103, bd Saint-Michel, 75005, Paris, France\\
\and
Institut d'Astrophysique Spatiale, CNRS (UMR8617) Universit\'{e} Paris-Sud 11, B\^{a}timent 121, Orsay, France\\
\and
Institut d'Astrophysique de Paris, CNRS (UMR7095), 98 bis Boulevard Arago, F-75014, Paris, France\\
\and
Institute for Space Sciences, Bucharest-Magurale, Romania\\
\and
Institute of Astronomy and Astrophysics, Academia Sinica, Taipei, Taiwan\\
\and
Institute of Astronomy, University of Cambridge, Madingley Road, Cambridge CB3 0HA, U.K.\\
\and
Institute of Theoretical Astrophysics, University of Oslo, Blindern, Oslo, Norway\\
\and
Instituto de Astrof\'{\i}sica de Canarias, C/V\'{\i}a L\'{a}ctea s/n, La Laguna, Tenerife, Spain\\
\and
Instituto de F\'{\i}sica de Cantabria (CSIC-Universidad de Cantabria), Avda. de los Castros s/n, Santander, Spain\\
\and
Jet Propulsion Laboratory, California Institute of Technology, 4800 Oak Grove Drive, Pasadena, California, U.S.A.\\
\and
Jodrell Bank Centre for Astrophysics, Alan Turing Building, School of Physics and Astronomy, The University of Manchester, Oxford Road, Manchester, M13 9PL, U.K.\\
\and
Kavli Institute for Cosmology Cambridge, Madingley Road, Cambridge, CB3 0HA, U.K.\\
\and
LAL, Universit\'{e} Paris-Sud, CNRS/IN2P3, Orsay, France\\
\and
LERMA, CNRS, Observatoire de Paris, 61 Avenue de l'Observatoire, Paris, France\\
\and
Laboratoire AIM, IRFU/Service d'Astrophysique - CEA/DSM - CNRS - Universit\'{e} Paris Diderot, B\^{a}t. 709, CEA-Saclay, F-91191 Gif-sur-Yvette Cedex, France\\
\and
Laboratoire Traitement et Communication de l'Information, CNRS (UMR 5141) and T\'{e}l\'{e}com ParisTech, 46 rue Barrault F-75634 Paris Cedex 13, France\\
\and
Laboratoire de Physique Subatomique et de Cosmologie, Universit\'{e} Joseph Fourier Grenoble I, CNRS/IN2P3, Institut National Polytechnique de Grenoble, 53 rue des Martyrs, 38026 Grenoble cedex, France\\
\and
Laboratoire de Physique Th\'{e}orique, Universit\'{e} Paris-Sud 11 \& CNRS, B\^{a}timent 210, 91405 Orsay, France\\
\and
Lawrence Berkeley National Laboratory, Berkeley, California, U.S.A.\\
\and
Max-Planck-Institut f\"{u}r Astrophysik, Karl-Schwarzschild-Str. 1, 85741 Garching, Germany\\
\and
Max-Planck-Institut f\"{u}r Extraterrestrische Physik, Giessenbachstra{\ss}e, 85748 Garching, Germany\\
\and
McGill Physics, Ernest Rutherford Physics Building, McGill University, 3600 rue University, Montr\'{e}al, QC, H3A 2T8, Canada\\
\and
MilliLab, VTT Technical Research Centre of Finland, Tietotie 3, Espoo, Finland\\
\and
Moscow Institute of Physics and Technology, Dolgoprudny, Institutsky per., 9, 141700, Russia\\
\and
Niels Bohr Institute, Blegdamsvej 17, Copenhagen, Denmark\\
\and
Observational Cosmology, Mail Stop 367-17, California Institute of Technology, Pasadena, CA, 91125, U.S.A.\\
\and
Optical Science Laboratory, University College London, Gower Street, London, U.K.\\
\and
SB-ITP-LPPC, EPFL, CH-1015, Lausanne, Switzerland\\
\and
SISSA, Astrophysics Sector, via Bonomea 265, 34136, Trieste, Italy\\
\and
SUPA, Institute for Astronomy, University of Edinburgh, Royal Observatory, Blackford Hill, Edinburgh EH9 3HJ, U.K.\\
\and
School of Physics and Astronomy, Cardiff University, Queens Buildings, The Parade, Cardiff, CF24 3AA, U.K.\\
\and
School of Physics and Astronomy, University of Nottingham, Nottingham NG7 2RD, U.K.\\
\and
Space Research Institute (IKI), Russian Academy of Sciences, Profsoyuznaya Str, 84/32, Moscow, 117997, Russia\\
\and
Space Sciences Laboratory, University of California, Berkeley, California, U.S.A.\\
\and
Special Astrophysical Observatory, Russian Academy of Sciences, Nizhnij Arkhyz, Zelenchukskiy region, Karachai-Cherkessian Republic, 369167, Russia\\
\and
Stanford University, Dept of Physics, Varian Physics Bldg, 382 Via Pueblo Mall, Stanford, California, U.S.A.\\
\and
Sub-Department of Astrophysics, University of Oxford, Keble Road, Oxford OX1 3RH, U.K.\\
\and
T\"{U}B\.{I}TAK National Observatory, Akdeniz University Campus, 07058, Antalya, Turkey\\
\and
Theory Division, PH-TH, CERN, CH-1211, Geneva 23, Switzerland\\
\and
UPMC Univ Paris 06, UMR7095, 98 bis Boulevard Arago, F-75014, Paris, France\\
\and
Universit\'{e} de Toulouse, UPS-OMP, IRAP, F-31028 Toulouse cedex 4, France\\
% ...
}

   \title{\Planck\ 2013 results. XX. Cosmology from
      Sunyaev--Zeldovich cluster counts}

    \abstract{We present constraints on cosmological parameters using
      number counts as a function of redshift for a sub-sample of 189
      galaxy clusters from the \Planck\ SZ (PSZ) catalogue. The PSZ is
      selected through the signature of the Sunyaev--Zeldovich (SZ)
      effect, and the sub-sample used here has a signal-to-noise
      threshold of seven, with each object confirmed as a cluster and
      all but one with a redshift estimate. We discuss the
      completeness of the sample and our construction of a likelihood
      analysis. Using a relation between mass $M$ and SZ signal $Y$
      calibrated to X-ray measurements, we derive constraints on the
      power spectrum amplitude $\sigma_8$ and matter density parameter
      $\Omega_{\mathrm{m}}$ in a flat $\Lambda$CDM model.  We test the
      robustness of our estimates and find that possible biases in the
      $Y$--$M$ relation and the halo mass function are larger than the
      statistical uncertainties from the cluster sample.  
        Assuming the X-ray determined mass to be biased low relative
        to the true mass by between zero and 30\%, motivated by
        comparison of the observed mass scaling relations to those
        from a set of numerical simulations, we find that
        $\sigma_8=0.75\pm 0.03$, $\Omega_{\mathrm{m}}=0.29\pm 0.02$,
        and $\sigma_8(\Omega_{\mathrm{m}}/0.27)^{0.3} = 0.764 \pm 0.025$.  The
        value of $\sigma_8$ is degenerate with the mass bias; if the
        latter is fixed to a value of 20\% (the central value from
        numerical simulations) we find
        $\sigma_8(\Omega_{\mathrm{m}}/0.27)^{0.3}=0.78\pm 0.01$ and a
        tighter one-dimensional range $\sigma_8=0.77\pm 0.02$.  We
      find that the larger values of $\sigma_8$ and
      $\Omega_{\mathrm{m}}$ preferred by \Planck's measurements of the
      primary CMB anisotropies can be accommodated by a mass bias of
      about  $40\%$. Alternatively, consistency with the primary
      CMB constraints can be achieved by inclusion of processes that
      suppress power on small scales relative to the $\Lambda$CDM
      model, such as a component of massive neutrinos. We place our
      results in the context of other determinations of cosmological
      parameters, and discuss issues that need to be resolved in order
      to make further progress in this field.}


\authorrunning{Planck Collaboration}
\titlerunning{Cosmology from SZ cluster counts}

   \maketitle

%\clearpage
%\allearlypapers

%________________________________________________________________

\begin{figure*}
\centering
\includegraphics[width=14cm]{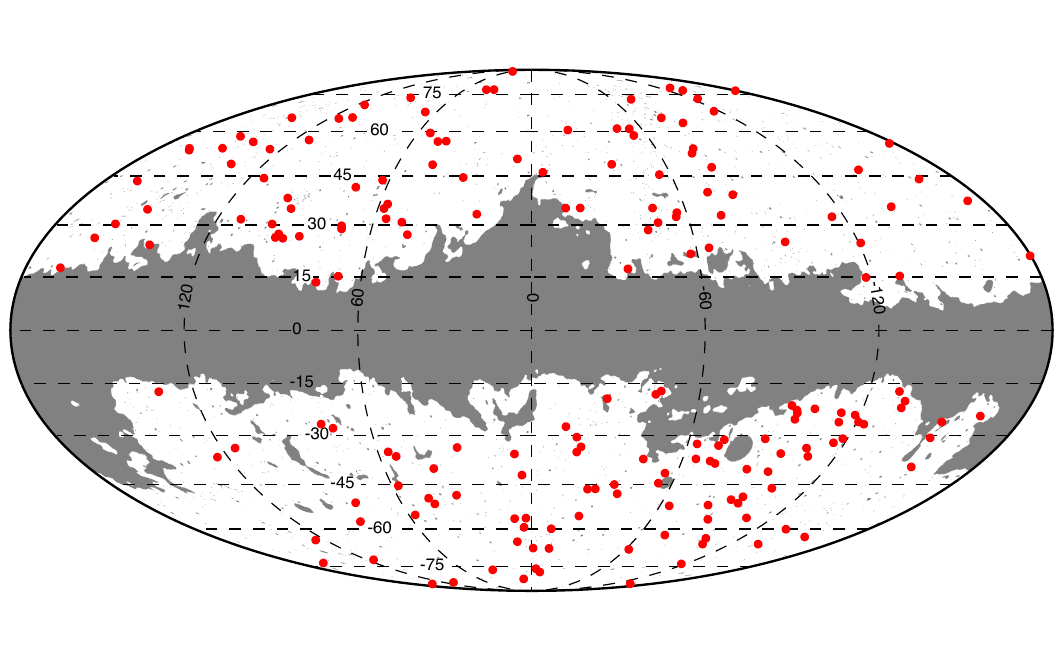}
\caption{Distribution on the sky of the \Planck\ SZ cluster
  sub-sample used in this paper, with the 35\% mask overlaid.}
\label{fig:sky}
\end{figure*}

%% INTRODUCTION
\section{Introduction}

This paper, one of a set associated with the 2013 release of data from
the \Planck\footnote{\Planck\ (\url{http://www.esa.int/Planck}) is a
  project of the European Space Agency (ESA) with instruments provided
  by two scientific consortia funded by ESA member states (in
  particular the lead countries France and Italy), with contributions
  from NASA (USA) and telescope reflectors provided by a collaboration
  between ESA and a scientific consortium led and funded by Denmark.}
mission \citep{planck2013-p01}\all2013resultspapers, describes the
constraints on cosmological parameters using number counts as a
function of redshift for a sample of 189 galaxy clusters.

Within the standard picture of structure formation, galaxies aggregate
into clusters of galaxies at late times, forming bound structures at
locations where the initial fluctuations create the deepest potential
wells. The study of these galaxy clusters has played a significant
role in the development of cosmology over many years (see, for
example, \citealt{1980ApJ...236..373P}, \citealt{1992A&A...262L..21O},
\citealt{1993Natur.366..429W}, \citealt{1996ApJ...462...32C},
\citealt{2005RvMP...77..207V}, \citealt{2009ApJ...691.1307H},
\citealt{vik09} and \citealt{Allen:2011zs}).  More recently, as
samples of clusters have increased in size and variety, number counts
inferred from tightly-selected surveys have been used to obtain
detailed constraints on the cosmological parameters.

The early galaxy cluster catalogues were constructed by eye from
photographic plates with a `richness' (or number of galaxies)
attributed to each cluster \citep{Abell1958,Abell1989}. As time has
passed, new approaches for selecting clusters have been developed, most
notably using X-ray emission due to thermal Bremsstrahlung radiation
from the hot gas that makes up most of the baryonic matter in the
cluster. X-ray cluster surveys include both the NORAS
\citep{Boehringer2000} and REFLEX \citep{Boehringer2004} surveys, based
on ROSAT satellite observations, which have been used as
source catalogues for higher-precision observations by the
\textit{Chandra} and \textit{XMM-Newton} satellites, as well as surveys with
\textit{XMM-Newton}, including the XMM Cluster Survey (XCS,
\citealt{Mehr12}) and the
XMM Large Scale Structure survey (XMM-LSS, \citealt{Willis13}).

To exploit clusters for cosmology, a key issue is how the properties
used to select and characterize the cluster are related to the total
mass of the cluster, since this is the quantity most readily predicted
using theoretical models.  Galaxies account for a small fraction of the
cluster mass, and the scatter between richness and mass appears to be
large.  However, there are a number of other possibilities. In
particular, there are strong correlations between the total mass and
both the integrated X-ray surface brightness and X-ray temperature,
making them excellent mass proxies.

The Sunyaev--Zeldovich (SZ) effect
\citep{1970Ap&SS...7...20S,1969Ap&SS...4..301Z} is the inverse Compton
scattering of cosmic microwave background (CMB) photons by the hot gas
along the line of sight, and this is most significant when the line of
sight passes through a galaxy cluster. It leads to a decrease in the
overall brightness temperature in the Rayleigh--Jeans portion of the
spectrum and an increase in the Wien tail, with a null around
$217\,{\mathrm{GHz}}$ (see \citealt{1999PhR...310...97B} for a
review). The amplitude of the SZ effect is given by the integrated
pressure of the gas within the cluster along the line of sight.
Evidence both from observation
\citep{2012ApJ...754..119M,planck2012-III} and from numerical
simulations
\citep{2001ApJ...562.1086S,2004MNRAS.348.1401D,2005ApJ...623L..63M,
  nag06,kay12} suggests that the SZ effect is an
excellent mass proxy.  A number of articles have discussed the
possibility of using SZ-selected cluster samples to constrain
cosmological parameters
\citep{bar96,agh97,2001ApJ...553..545H,2001ApJ...560L.111H,
  2002PhRvL..88w1301W,2002MNRAS.331..556D,2003PhRvD..68h3506B}.

This paper describes the constraints on cosmological parameters
imposed by a high signal-to-noise (S/N) sub-sample of the \Planck\ SZ
Catalogue (PSZ, see \citealt{planck2013-p05a}, henceforth Paper 1, for
details of the entire catalogue) containing nearly 200 clusters (shown in
Fig.~\ref{fig:sky}). This sub-sample has been selected to be pure, in
the sense that all the objects within it have been confirmed as
clusters via additional observations, either from the literature or
undertaken by the Planck collaboration.  In addition all objects but
one have a measured redshift, either photometric or
spectroscopic. This is the largest SZ-selected sample of clusters used
to date for this purpose.  We will show that it is the systematic
uncertainties from our imperfect knowledge of cluster properties that
dominate the overall uncertainty on cosmological constraints.

The \Planck\ cluster sample is complementary to those from
observations using the South Pole Telescope (SPT, \citealt{SPT}) and
the Atacama Cosmology Telescope (ACT, \citealt{ACT}), whose teams
recently published the first large samples of SZ-selected clusters
\citep{Reichardt2012,Hass13}. The resolution of \Planck\ at the
relevant frequencies is between 5 and 10 arcmin, whereas that for ACT
and SPT is about $1\,{\mathrm{arcmin}}$, but the \Planck\ sky
coverage is much greater. This means that \Planck\ typically finds
larger, more massive, and lower-redshift clusters than those found by SPT and ACT.

Our strategy is to focus on number counts of
clusters, as a function of redshift, above a high S/N threshold of seven
and to explore the robustness of the results. We do not use the
observed SZ brightness of the clusters, due to the significant
uncertainty caused by the size--flux degeneracy as discussed in
Paper~1. Accordingly, our theoretical modelling of the cluster
population is directed only at determining the expected number of clusters
in each redshift bin exceeding the S/N threshold. The predicted and
observed numbers of clusters are then compared in order to obtain the
likelihood. In the future, we will make use of the SZ-estimated mass
and a larger cluster sample to extend the analysis to broader
cosmological scenarios.

This paper is laid out as follows. We describe the theoretical
modelling of the redshift number counts in Sect.~\ref{sec:modelling},
while Sect.~\ref{sec:sampleselfunc} presents the PSZ
cosmological sample and selection function used in this work.  The
likelihood we adopt for putting constraints on cosmological parameters
is given in Sect.~\ref{sec:likelihood}. Section~\ref{sec:constraints}
presents our results on cosmological parameter estimation and assesses their
robustness. We discuss how they fit in with other
cluster and cosmological constraints in Sect.~\ref{sec:discussion},
before providing a final summary. A detailed discussion of our
calibration of the SZ flux versus mass relation and its uncertainties
is given in Appendix A.

\section{Modelling cluster number counts \label{sec:modelling}}
 
\subsection{Model definitions}

We parameterize the standard cosmological model as follows. The
densities of various components are specified relative to the
present-day critical density, with
$\Omega_{\mathrm{X}}=\rho_{\mathrm{X}}/\rho_{\mathrm{crit}}$ denoting
that for component X. These components always include matter,
$\Omega_{\mathrm{m}}$, and a cosmological constant $\Omega_{\Lambda}$.
For this work we assume that the Universe is flat, that is,
$\Omega_{\mathrm{m}}+\Omega_{\Lambda}=1$, and the optical depth to
reionization is fixed at $\tau=0.085$ except in the CMB+SZ analyses.
The present-day expansion rate of the Universe is quantified by the
Hubble constant $H_0=100\,h\,{\mathrm{km}}\,{\mathrm{
    s}}^{-1}\,{\mathrm{Mpc}}^{-1}$.

The cluster number counts are very sensitive to the amplitude of the
matter power spectrum. When studying cluster counts it is usual to
parametrize this in terms of the density variance in spheres of radius
$8 h^{-1} \, \mathrm{Mpc}$, denoted $\sigma_8$, rather than overall
power spectrum amplitude, $A_{\mathrm{s}}$. In cases where we include
primary CMB data we use $A_{\mathrm{s}}$ and compute $\sigma_8$
as a derived parameter. In addition to the parameters above, we 
allow the other standard cosmological parameters to vary:
$n_{\mathrm{s}}$ representing the spectral index of density
fluctuations; and $\Omega_{\mathrm{b}}h^2$ quantifying the baryon
density.

The number of clusters predicted to be observed by a survey in a given
redshift interval $[z_i, z_{i+1}]$ can be written
\begin{equation}
n_i = \int_{z_i}^{z_{i+1}} dz {dN \over dz},
\end{equation}
with 
\begin{equation}
\label{eq:dndz}
{dN \over dz} = \int d\Omega \int dM_{500} \,
\hat{\chi}(z,M_{500},l,b) \, {dN \over dz \, dM_{500} \,d\Omega}\,,
\end{equation}
where $d\Omega$ is the solid angle element and $M_{500}$ is the mass
within the radius where the mean enclosed density is 500 times the
critical density. The quantity $\hat{\chi}(z,M_{500},l,b)$ is the
survey completeness at a given location $(l,b)$ on the sky, given by
\begin{equation}
\hat{\chi} = \int dY_{500} \int d\theta_{500}
P(z,M_{500}|Y_{500},\theta_{500}) \, \chi(Y_{500},\theta_{500},l,b)\,.  
\end{equation}
Here $P(z,M_{500}|Y_{500},\theta_{500})$ is the distribution of
$(z,M_{500})$ for a given $(Y_{500},\theta_{500})$, where $Y_{500}$
and $\theta_{500}$ are the SZ flux and size of a cluster of redshift
and mass $(z,M_{500})$.

This distribution is obtained from the scaling relations between
$Y_{500}$, $\theta_{500}$, and $M_{500}$, discussed later in this
section.  Note that $\hat{\chi}(z,M_{500},l,b)$ depends on
cosmological parameters through $P(z,M_{500} | Y_{500},\theta_{500})$,
while the completeness in terms of the observables,
$\chi(Y_{500},\theta_{500},l,b)$, does not depend on the cosmology as
it refers directly to the observed quantities.

\begin{figure}
\centering
\includegraphics[width=8.8cm]{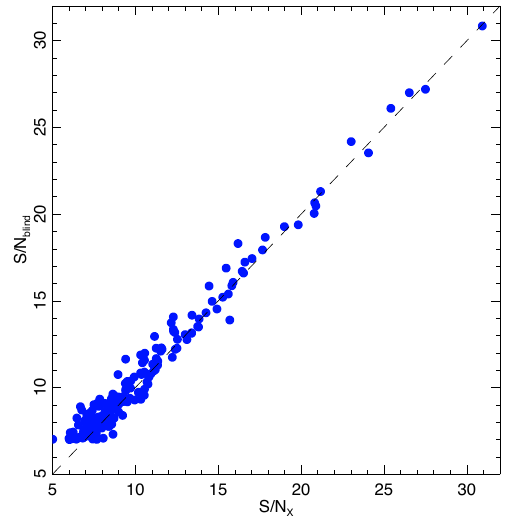}
\caption{Blind S/N versus S/N re-extracted at the X-ray position using
  the X-ray size, for the \texttt{MMF3} detections of \Planck\
  clusters that are associated with known X-ray clusters in the
  reference cosmological sample.  In contrast to the blind SZ flux,
  the blind S/N is in good agreement with S/N measured using X-ray
  priors. }
\label{fig:snr}
\end{figure}

For the present work, we restrict our analysis to the quantity $dN/dz$
that measures the total counts in redshift bins. In particular, we do
not use the blind SZ flux estimated by the cluster candidate
extraction methods that, as detailed in \citet{planck2011-5.1a}, is
found to be significantly higher than the flux predicted from X-ray
measurements.  In contrast to the blind SZ flux, the blind S/N is in
good agreement with the S/N measured using X-ray
priors. Figure~\ref{fig:snr} shows the blind S/N
(S/N$_{\mathrm{blind}}$) versus the S/N re-extracted at the X-ray
position and using the X-ray size (S/N$_{\mathrm{X}}$). The clusters
follow the equality line. In Sect.~\ref{sec:sampleselfunc}, we use the
(S/N$_{\mathrm{blind}}$) values to define our cosmological sample, while for
the predicted counts (defined in Sect.~\ref{sec:modelling}) we use the
completeness based on S/N$_{\mathrm{X}}$. Our analysis relies on the
good match between these two quantities.\footnote{The two signal-to-noises are actually estimated at two different positions on the sky (blind SZ and X-ray position), leading to different values of both the signal and the noise. It thus happens that the recomputed $S/N$ is higher than the blind SZ.}

To carry out a prediction of the counts expected in a survey, given
cosmological assumptions, we therefore need the following inputs:
\begin{itemize}
\item a mass function that tells us the number distribution of
  clusters with mass and redshift;
\item scaling relations that can predict observable quantities from
  the mass and redshift;
\item the completeness of the survey in terms of those observables,
  which tells us the probability that a model cluster would make it
  into the survey catalogue.
\end{itemize}
These are described in the remainder of this section and in the next.

\subsection{Mass function}

Our main results use the mass function from
\citet{2008ApJ...688..709T}, giving the number of haloes per unit volume:
\begin{equation}
\label{eq:mf}
{dN\over dM_{500}}(M_{500},z)=f(\sigma)\,{\rho_{\mathrm
  m}(z=0) \over M_{500}}\,{d\ln\sigma^{-1}\ \over dM_{500}} \,,
\end{equation}
where 
\begin{equation}
f(\sigma)=A\left[1+\left({\sigma\over
    b}\right)^{-a}\right]\exp\left(-{c\over\sigma^2}\right)\,, 
\end{equation}
and ${\rho}_\mathrm{m}(z=0)$ is the mean matter density at $z=0$. The
coefficients $A$, $a$, $b$ and $c$ are tabulated in
\citet{2008ApJ...688..709T} for different overdensities,
$\Delta_{\mathrm{mean}}$, with respect to the mean cosmic density, and
depend on $z$. Here we use $\Delta_\mathrm{critical}=500$
relative to the critical density, so we compute the relevant mass
function coefficients by interpolating the \citet{2008ApJ...688..709T}
tables for haloes with $\Delta_\mathrm{mean} \equiv
\Delta_\mathrm{critical}/\Omega_\mathrm{m}(z)=500/\Omega_\mathrm{m}(z)$,
where $\Omega_\mathrm{m}(z)$ is the matter density parameter at
redshift $z$.

The quantity $\sigma$ is the standard deviation, computed in linear
perturbation theory, of the density perturbations in a sphere of
radius $R$, which is related to the mass by $M=4\pi\rho_{\mathrm
  m}(z=0)R^3/3$. It is given by 
\begin{equation}
\sigma^2=\frac{1}{2\pi^2} \int dk \, k^2 P(k,z) |W(kR)|^2\,,
\end{equation}
where $P(k,z)$ is the matter power spectrum at redshift $z$, which we
compute for any given set of cosmological parameters using {\tt CAMB}
\citep{Lewis99}, and $W(x)=3(\sin x - x\cos x)/x^3$ is the filter
function of a spherical top hat of radius $R$.

The quantity $dN/(dz \, dM_{500} \, d\Omega)$ in Eq.~\ref{eq:dndz}
is computed by multiplying the mass function $dN(M_{500},z)/dM_{500}$
by the volume element $dV/(dz \, d\Omega)$.

As a baseline we use, except where stated otherwise, the
\citet{2008ApJ...688..709T} mass function, but we consider an
alternative mass function as a cross-check.  In a recent publication
by \citet{wat12}, a new mass function is extracted from the
combination of large cosmological simulations (typical particle
numbers of $3000^3$ to $6000^3$) with a very large dynamic range (size
from $11\,h^{-1}$ to $6000\, h^{-1}$Mpc), which extends the maximum
volume probed by Tinker et al.\ by two orders of magnitude.  The two
mass functions agree fairly well, except in the case of the most
massive objects, where Tinker et al.'s mass function predicts more
clusters than Watson et al.'s. The Tinker et al.\ mass function might
be derived from volumes that are not large enough to properly sample
the rarer clusters. These rare clusters are more relevant for
\Planck\ than for ground-based SZ experiments, which probe smaller
areas of the sky. The Watson et al.\ mass function is used only in
Sect.~\ref{sec:robust2}, which deals with mass function uncertainties.

\subsection{Scaling relations\label{sec:scaling}}

A key issue is to relate the observed SZ flux, $Y_{500}$, to the mass
$M_{500}$ of the cluster. As we show in Sect.~\ref{sec:constraints},
cosmological constraints are sensitive to the normalization, slope and scatter
of the assumed $Y_{500}$--$M_{500}$ relation. We thus paid
considerable attention to deriving the most accurate scaling relations
possible, with careful handling of statistical and systematic
uncertainties, and to testing their impact on the derived cosmological
parameters.

The baseline relation is obtained from an observational calibration of
the $Y_{500}$--$M_{500}$ relation on one-third of the cosmological
sample.  The calibration uses $M_{500}^{\YX}$, the mass derived from
the X-ray $\YX$--$M_{500}$ relation, as a mass proxy. Here $\YX$ is
the X-ray analogue of the SZ signal introduced by \citet{kra06}, as
defined in Appendix~\ref{app:scaling}.  $Y_{500}$ is then measured
interior to $R_{500}^{\YX}$, the radius corresponding to
$M_{500}^{\YX}$. The mean bias between $M_{500}^{\YX}$ and the true
mass, $(1-b)$, is assumed to account for all possible observational
biases (departure from hydrostatic equilibrium (HE), absolute
instrument calibration, temperature inhomogeneities, residual
selection bias, etc) as discussed in full in
Appendix~\ref{app:scaling}.  In practice, the plausible range for this
mean bias $(1-b)$ was estimated by comparing the observed
relation with predictions from several sets of numerical simulations,
as detailed in Appendix~\ref{app:scaling}.

The large uncertainties on $(1-b)$ are due to the dispersion in
predictions from the various simulation sets.  This is a major
  factor limiting the accuracy of our analysis. A value $(1-b)=0.8$
  could be considered as a best guess given available simulations,
  with no clear dependence on mass or redshift. From one cluster to
the next the ratio of $M^{Y_X}_{500}$ to the true mass is expected to be
stochastic, contributing to the scatter in the \mbox{$Y_{500}$--$M_{500}$}
relation given below.  A conspiracy of all possible sources of
  bias (departure from HE, absolute instrument calibration,
  temperature inhomogeneities, residual selection bias) would seem
  necessary to lead to a significantly lower value of $(1-b)$. This
  apparently implausible possibility needs to be excluded through
  tests using other probes such as baryon and gas fractions, gas
  pressure, etc.  As a baseline we take $(1-b)$ to vary within the
  range $[0.7,1.0]$ with a flat prior. We also consider, when
  analysing systematic uncertainties on the derived cosmological
  parameters, a case where the bias is fixed to the value
  $(1-b)=0.8$.

As detailed in Appendix~\ref{app:scaling}, we derive a baseline
relation for the mean SZ signal $\bar{Y}_{500}$ from a cluster of
given mass and redshift in the form
\begin{equation}
E^{-\beta}(z)\left[\frac{D_{\mathrm{ A}}^2(z) \, {\bar Y}_{500}}
  {\mathrm{ 10^{-4}\,Mpc^2}}\right] =  Y_\ast \left[ {h \over 0.7}
  \right]^{-2+\alpha} \left[\frac{(1-b)\,
    \Mv}{6\times10^{14}\,\mathrm{M}_\odot}\right]^{\alpha}\,,\label{scaling} 
\end{equation}
where $D_\mathrm{A}(z)$ is the angular-diameter distance to redshift
$z$ and $E^2(z)=\Omega_{\mathrm m}(1+z)^3+\Omega_{\Lambda}$. The
coefficients $Y_\ast$, $\alpha$, and $\beta$ are given in
Table~\ref{table:scaling}.

\begin{table}[tmb]                 
\begingroup
\newdimen\tblskip \tblskip=5pt
\caption{Summary of scaling-law parameters and error budget.}
  \label{table:scaling}
\nointerlineskip
\vskip -3mm
\footnotesize
\setbox\tablebox=\vbox{
   \newdimen\digitwidth 
   \setbox0=\hbox{\rm 0} 
   \digitwidth=\wd0 
   \catcode`*=\active 
   \def*{\kern\digitwidth}
   \newdimen\signwidth 
   \setbox0=\hbox{+} 
   \signwidth=\wd0 
   \catcode`!=\active 
   \def!{\kern\signwidth}
\halign{\hfil#\hfil\tabskip=2em & \hfil#\tabskip=0pt\cr                           % Template goes here.
\noalign{\doubleline}
                                    % Table headings go here.
%\noalign{\vskip 3pt\hrule\vskip 5pt}
Parameter & Value~~~~ \cr
\noalign{\vskip 3pt\hrule\vskip 1pt}
$\log Y_{\ast}$& $-0.19 \pm 0.02$ \cr
$\alpha$ & $1.79\pm0.08$ \cr
$\beta$ & $0.66\pm 0.50$  \cr
$\sigma_{\log Y}$ &$0.075 \pm 0.01$ \cr                            % Body of table goes here.
\noalign{\vskip 1pt\hrule\vskip 1pt}}}
\endPlancktable                    % ends one-column \halign
%\endPlancktablewide                 % ends two-column \halign
\endgroup
\tablefoot{$\beta$ is kept fixed at its central value except in 
Sect.~\ref{sec:robust2}.}
\end{table}                        % table* is a two-column table.  Drop the * for one column.

Equation~\ref{scaling} has an estimated intrinsic
scatter\footnote{Throughout this article, $\log$ is base 10 and $\ln$
  is base $e$.}  $\sigma_{\log Y}=0.075$, which we take to be
independent of redshift (see Appendix~\ref{app:scaling}). This is
incorporated by drawing the cluster's $Y_{500}$ from a log-normal
distribution
\begin{equation}
{\cal P}(\log Y_{500})={1\over
  \sqrt{2\pi\sigma_{\log Y}^2}}\exp\left[{-\frac{\log^2(Y_{500}/{\bar
      Y}_{500})}{ 2\sigma_{\log Y}^2}}\right]\,, 
\label{eq:scatter}
\end{equation}
where ${\bar Y}_{500}$ is given by Eq.~\ref{scaling}. Inclusion of
this scatter increases the number of clusters expected at a given S/N;
since the cluster counts are a steep function of $M_{500}$ in the
range of mass in question, there are more clusters that scatter
upwards from below the limit given by the zero-scatter scaling
relation than those that scatter downwards.

In addition to Eq.~\ref{scaling} we need a relation between
$\theta_{500}$ (in fact $\theta_{500}^{\YX}$, the angular size
corresponding to the physical size $R_{500}^{\YX}$), the aperture used
to extract $Y_{500}$, and $M_{500}$. Since $M_{500}=500 \times 4\pi
\rho_\mathrm{crit} R_{500}^3/3$ and
$\theta_{500}=R_{500}/D_\mathrm{A}$, this can be expressed as
\begin{equation}
{\bar \theta_{500}}=\theta_\ast
\left[\frac{h}{0.7}\right]^{-2/3}\left[{(1-b)\,M_{500}\over 3\times
    10^{14} \msol}\right]^{1/3} \,E^{-2/3}(z)\,\left[{D_{\mathrm
      A}(z)\over 500\,{\mathrm{Mpc}}}\right]^{-1}\!\!, 
\label{angle}
\end{equation}
where $\theta_\ast=6.997 \, \mathrm{arcmin}$.

\subsection{Limiting mass}

\begin{figure}
\centering
\includegraphics[width=8.8cm]{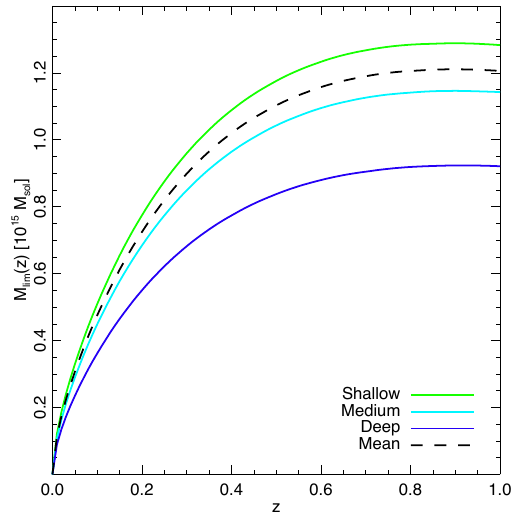}
\caption{ Limiting mass as a function of $z$ for the selection
  function and noise level computed for three zones (deep, blue;
  medium, cyan; shallow, green), and on average for the unmasked sky
  (dashed black).}
\label{fig:mlim}
\end{figure}

One can use Eqs.~\ref{scaling} and \ref{angle} to compute the limiting
mass at a point on the sky where the noise level, $\sigma_{Y}$, has
been computed as described in Sect.~\ref{sec:sampleselfunc}. As the
latter is not homogeneous on the sky, we show in Fig.~\ref{fig:mlim}
the limiting mass, defined at 50\% completeness, as a function of
redshift for three different zones, deep, medium, and shallow, covering
respectively, 3.5\%, 47.8\%, and 48.7\% of the unmasked sky. For each
line a S/N cut of 7 has been adopted.

\subsection{Implementation}

We have implemented three independent versions of the computation of
counts and constraints. The differences in predicted counts are of the
order of a few percent, which translates to less than a tenth of $1\,
\sigma$ on the cosmological parameters of interest.

\section{The \Planck\ cosmological samples \label{sec:sampleselfunc}}

\subsection{Sample definition}
\label{sec:sample:def}

The reference cosmological sample is constructed from the PSZ
Catalogue published in \citet{planck2013-p05a} and made public
with the first release of \Planck\ cosmological products. It is based
on the SZ detections performed with the matched multi-filter (MMF)
method \texttt{MMF3} \citep{mel06}, which relies on the use of a filter of
adjustable width $\theta_{500}$ chosen to maximize the S/N of the
detection. In order to ensure a high purity and to maximize the number
of redshifts, the cosmological sample was constructed by selecting the
SZ detections above a S/N threshold of 7 outside Galactic and point
source masks covering 35\% of the sky, as discussed in Paper 1.  From
the original PSZ, only the information on S/N (for the selection) and
redshift are used.

This sample contains 189 candidates. All but one are confirmed
bona fide clusters with measured redshifts, including 184
spectroscopic redshifts.  Among these confirmed clusters 12 were
confirmed with follow-up programmes conducted by the
Planck collaboration (see Paper 1 for details). The remaining
non-confirmed cluster candidate is a high-reliability {\sc class1}
candidate, meaning that its characterization as a cluster is supported
by data in other wavebands (see Paper 1 for details).  It is thus
considered as a bona fide cluster. The distribution on the
sky of this baseline cosmological sample is shown in
Fig.~\ref{fig:sky}.

In addition to our reference sample, we consider two other samples
drawn from the PSZ for consistency checks. One is based on the
detections from the second implementation of the MMF algorithm,
\texttt{MMF1}, described in Paper 1. It contains 188 clusters with S/N
$>7$ and no missing redshifts, with almost complete overlap with the
baseline sample (187 clusters in common). The third sample considered
in the present study is also based on \texttt{MMF3} detections but
with a higher S/N cut of S/N $>8$. It allows us to test selection
effects and to probe the consistency of the results as a function of
the S/N cut. It contains 136 clusters, all with measured redshifts.

The selection function for each of these samples is constructed as
described in the next section.  

\subsection{Completeness}

The completeness of the reference cosmological sample is computed with
two distinct and complementary approaches: a semi-analytic approach
based on the assumption of Gaussian uncertainties, and a computational
approach based on Monte Carlo cluster injection into real sky maps.

The completeness $\chi$ can be evaluated analytically by setting the
probability of the measured SZ flux, $Y_{500}$, to be Gaussian
distributed with a standard deviation equal to the noise,
$\sigma_{Y_{500}}(\theta_{500}, l, b)$, computed for each size
$\theta_{500}$ of the MMF filter and at each position $(l,b)$ on the
sky:
\begin{equation}
  \chi_{\mathrm{erf}}(Y_{500}, \theta_{500}, l, b) = \frac{1}{2}\left[ {1 +
      {\mathrm{erf}} \left( {{Y_{500} - X \,
            \sigma_{Y_{500}}(\theta_{500}, l, b) } \over  {{\sqrt 2}
            \, \sigma_{Y_{500}}(\theta_{500}, l, b)}} \right) } \right]  \,,
\label{eq:completness-chi}
\end{equation}
where $X=7$ is the S/N threshold and the error function is defined as
usual by
\begin{equation}
{\mathrm{erf}}(u)= {2 \over \sqrt \pi} \int_0^u \exp \left(-t^2\right)
dt\,. 
\end{equation}
$\chi_{\mathrm{erf}}(Y_{500},\theta_{500},l,b)$ thus lies in the range
$[0,1]$ and gives the probability for a cluster of flux $Y_{500}$ and
size $\theta_{500}$ at position $(l,b)$ to be detected at S/N $\ge X$.

\begin{figure}
\centering
\includegraphics[width=8.8cm]{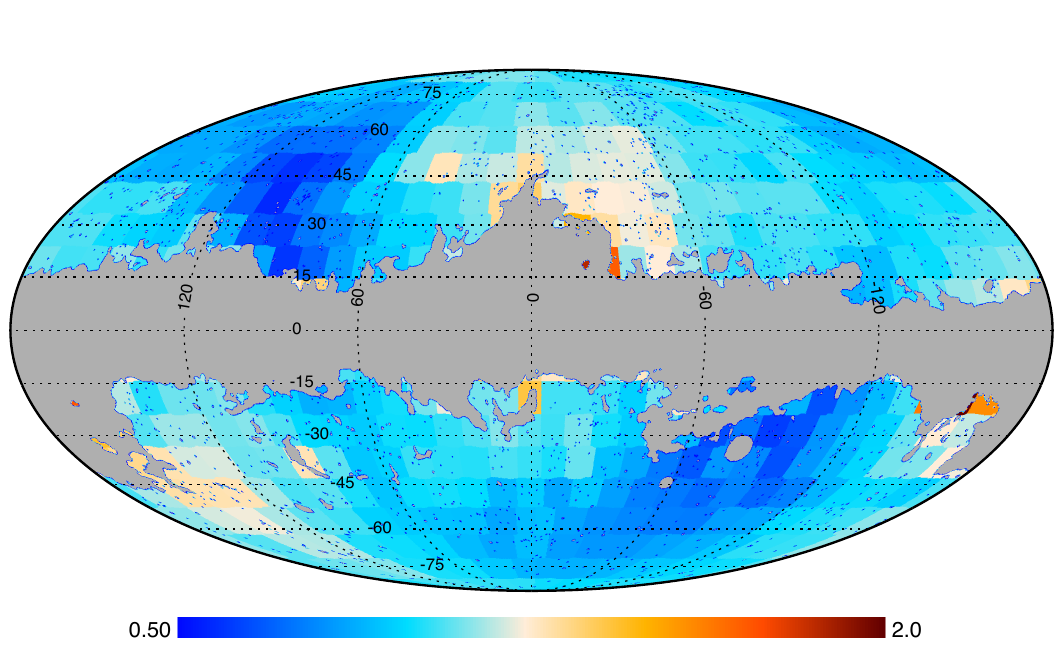}
\caption{Noise map $\sigma_{Y_{500}}(\theta_{500})$ for
  $\theta_{500}=6\; \mathrm{arcmin}$. The PSZ is limited by
  instrumental noise at high ($|b|>20\deg$) Galactic latitude (deeper
  at ecliptic poles) and foreground noise at low Galactic
  latitude. The scale of the map ranges from 0.5 to 2 times the mean
  noise of the map, which is $\langle \sigma_{Y_{500}}(6
    \;\mathrm{arcmin})\rangle=2.2\times 10^{-4} \mathrm{arcmin}^2$.}
\label{fig:noisemaps}
\end{figure}

The noise estimate $\sigma_{Y_{500}}(\theta_{500}, l, b)$ is a
by-product of the detection algorithm and can be written in the form
(see e.g., \citealt{mel06})
\begin{equation}
\sigma_{Y_{500}}(\theta_{500}, l, b) = \left [\int d^2k \;
  \vec{F}_{\theta_{500}}^t(\vec{k}) \cdot \vec{P}^{-1}(\vec{k},l,b)
  \cdot  \vec{F}_{\theta_{500}}(\vec{k}) \right ]^{-1/2} \,,
\end{equation}
with $\vec{F}_{\theta_{500}}(\vec{k})$ being a vector of dimension
$N_{\mathrm{freq}}$ (the six highest \planck\ frequencies here)
containing the beam-convolved cluster template scaled to the known SZ
frequency dependence. The cluster template assumed is the non-standard
universal pressure profile from
\citet{arn10}. $\vec{P}(\vec{k},l,b)$ is the noise power spectrum
(dimension $N_{\mathrm{freq}} \times N_{\mathrm{freq}}$) directly
estimated from the data at position $(l,b)$.
Figure~\ref{fig:noisemaps} shows $\sigma_{Y_{500}}(\theta_{500}, l,
b)$ for $\theta_{500}=6 \, {\mathrm{arcmin}}$ in a Mollweide
projection with the Galactic mask used in the analysis applied. As
expected, the noise at high Galactic latitude is lower than in the
areas contaminated by diffuse Galactic emission. The ecliptic pole
regions have the lowest noise level, reflecting the longer
\planck\ integration time in these high-redundancy areas.

\begin{figure}
\centering
\includegraphics[width=8.8cm]{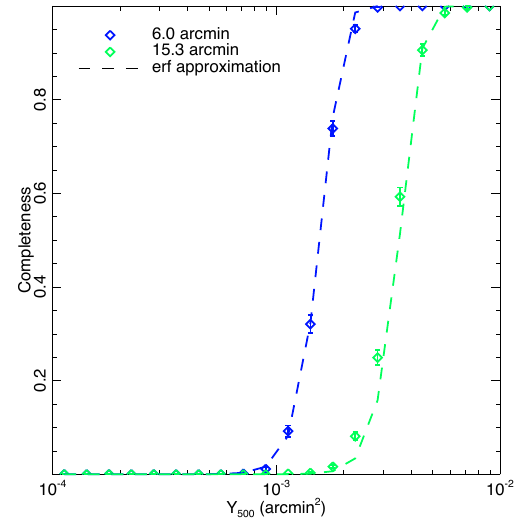}
\caption{Completeness averaged over the unmasked sky as a function of
  $Y_{500}$ for two different filter sizes, $\theta_{500}=6$ and
  $15.3$ arcmin. The dashed lines show the semi-analytic approximation
  of Eq.~\ref{eq:completness-chi}, while the diamonds with errors show the completeness estimated by the MC injection technique.}
\label{fig:erf}
\end{figure}

The Monte Carlo (MC) completeness is calculated by injecting simulated
clusters into real sky maps following the method presented in Paper 1,
with the modifications that the 65\% Galactic dust mask and a S/N
$>7$ threshold are applied to match the cosmological sample
definition.  The Monte Carlo completeness encodes effects not probed
by the erf approximation, including the variation of cluster pressure
profiles around the fiducial pressure profile used in the MMF,
spatially-varying and asymmetric effective beams, and the effects of
correlated non-Gaussian uncertainties in the estimation of
$(Y_{500},\theta_{500})$.  As shown in Fig.~\ref{fig:erf}, the
erf-based formula for the completeness is a good approximation to the
Monte Carlo completeness. The agreement is best for the typical sizes
probed by \Planck\ (5 to 10 arcmin), though the two determinations of
the completeness start to deviate for small and large sizes, due to
beam and profile effects, respectively.  For simplicity, we chose the
erf formulation as the baseline.  The effect of using the Monte Carlo
completeness instead is discussed in Sect.~\ref{sec:errorbudget}.

\begin{table*}[tmb]                 % table* is a two-column table.  Drop the * for one column.
\begingroup
\newdimen\tblskip \tblskip=5pt
\caption{\footnotesize{Best-fit cosmological parameters for
    various combinations of data and analysis methods. } }
\label{tab:constraints}
\nointerlineskip
\vskip -3mm
\footnotesize
\setbox\tablebox=\vbox{
   \newdimen\digitwidth 
   \setbox0=\hbox{\rm 0} 
   \digitwidth=\wd0 
   \catcode`*=\active 
   \def*{\kern\digitwidth}
   \newdimen\signwidth 
   \setbox0=\hbox{+} 
   \signwidth=\wd0 
   \catcode`!=\active 
   \def!{\kern\signwidth}
\halign{#\hfil\tabskip=2em & \hfil#\hfil & \hfil#\hfil & \hfil#\hfil & \hfil#\hfil\tabskip=0pt\cr
\noalign{\doubleline}
Data  & $\sigma_8 (\Omega_\mathrm{m}/0.27)^{0.3}$ &  $\Omega_\mathrm{m}$
& $\sigma_8$ &  $1-b$    \cr
\noalign{\vskip 3pt\hrule\vskip 5pt}
 \Planck\ SZ +BAO+BBN  & $ 0.764 \pm 0.025 $ & $
 0.29 \pm 0.02 $ & $ 0.75 \pm 0.03 $ &  [0.7,1]   \cr
 \Planck\ SZ +HST+BBN  & $ 0.774 \pm 0.024 $ & $
 0.28 \pm 0.03 $ & $ 0.77 \pm 0.03 $ &  [0.7,1]   \cr \noalign{\vskip 3pt\hrule\vskip 5pt}
 \Planck\ SZ +BAO+BBN    & $ 0.782 \pm 0.010 $ & $ 0.29 \pm 0.02 $ &
 $ 0.77 \pm 0.02 $ & $0.8$ \cr 
 \texttt{MMF1} sample +BAO+BBN       & $0.800\pm 0.010$ &  $0.29 \pm 0.02$ &
 $0.78 \pm 0.02$ &  $0.8$    \cr
 \texttt{MMF3} S/N $>8$ +BAO+BBN      & $ 0.785 \pm 0.011 $ & $ 0.29 \pm 0.02 $
 & $ 0.77 \pm 0.02 $ & $0.8$ \cr   
 \Planck\ SZ +BAO+BBN (MC completeness)      & $ 0.778 \pm 0.010 $ & $
 0.30 \pm 0.03 $ & $ 0.75 \pm 0.02 $ & $0.8$ \cr
 \Planck\ SZ +BAO+BBN (Watson et al.\ mass function)      & $0.802\pm0.014$ &
 $0.30\pm 0.01$ 
 & $0.77\pm0.02$ &  $0.8$  \cr
\noalign{\vskip 5pt\hrule\vskip 3pt}}}
%\endPlancktable                    % ends one-column \halign
\endPlancktablewide                 % ends two-column \halign
%\tablenote a Footnote a.\par
%\tablenote b Footnote b.\par
\endgroup
\tablefoot{For
    the analysis using the Watson et al.\ mass function, or $(1-b)$ in
    $[0.7,1]$, the degeneracy line is different, and thus the value of
    $\sigma_8 (\Omega_\mathrm{m}/0.27)^{0.3}$ is just illustrative.}
\end{table*}                        % table* is a two-column table.  Drop the * 

\section{Likelihood and Markov chain Monte Carlo \label{sec:likelihood}}

\subsection{The likelihood}

We now have all the information needed to predict the counts in
redshift bins for our theoretical models.  To obtain cosmological
constraints with the PSZ sample presented in
Sect.~\ref{sec:sampleselfunc}, we construct a likelihood function
based on Poisson statistics \citep{cash79}:
\begin{equation} 
\ln L = \ln \mathcal{P} (N_i|n_i) = \sum_{i=1}^{N_\mathrm{b}} [N_i \ln(n_i) -
  n_i - \ln(N_i!)]\,,
\end{equation}
where $\mathcal{P}(N_i|n_i)$ is the probability of finding $N_i$
clusters in each of $N_\mathrm{b}$ bins given an expected number of
$n_i$ in each bin in redshift.  The likelihood includes bins that
contain no observed clusters. As a baseline, we assume bins in
redshift of $\Delta z = 0.1$ and we checked that our results are
robust when changing the bin size between $0.05$ and $0.2$.  The
modelled expected number $n_i$ depends on the bin range in redshift
and on the cosmological parameters, as described in
Sect.~\ref{sec:modelling}. It also depends on the scaling relations
and the selection function of the observed sample.  The parameters of
the scaling relations between flux (or size) and mass and redshift are
taken to be Gaussian distributed with central values and uncertainties
stated in Table~\ref{table:scaling}, and with the scatter in $Y_{500}$
incorporated into the method via the log-normal distribution with width $\sigma_{\log
  Y}$.

\begin{figure*}
\centering
\includegraphics[width=16cm]{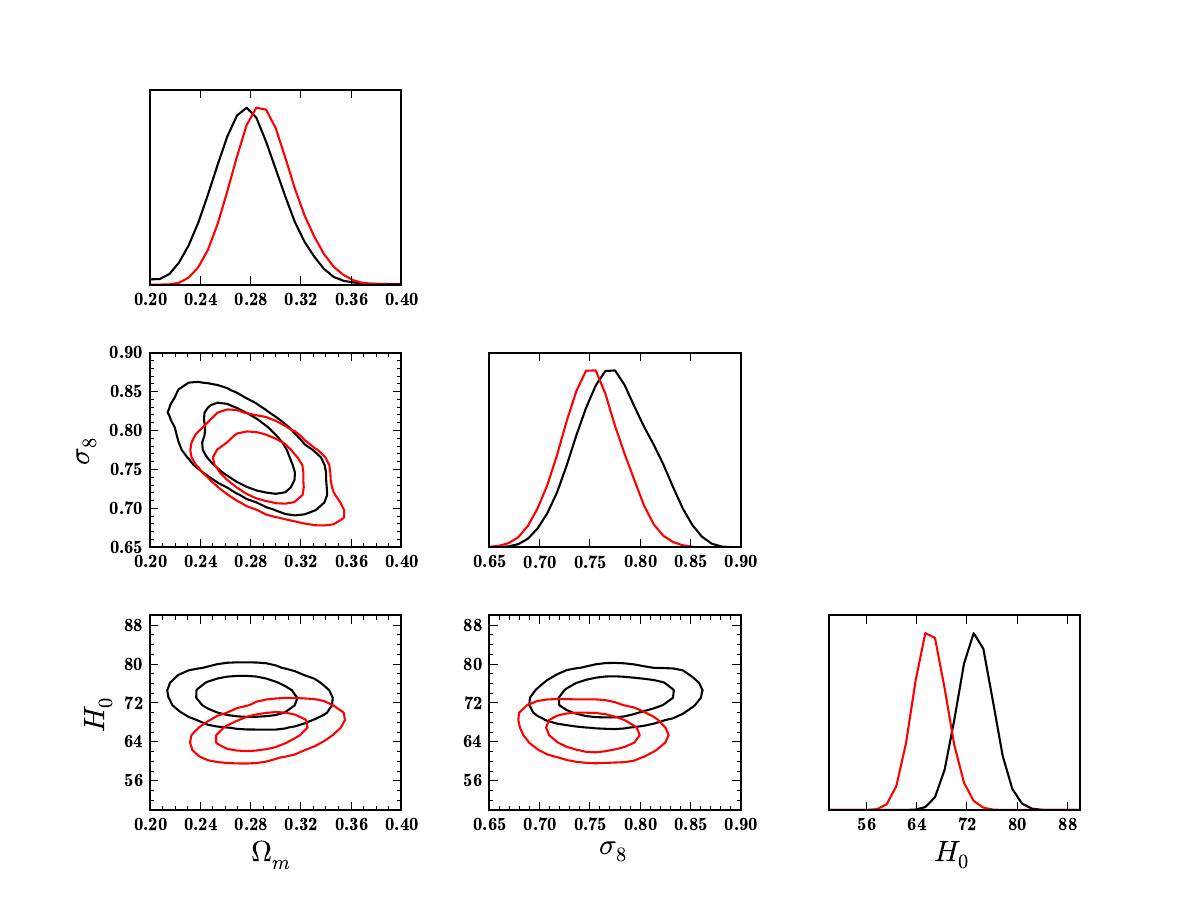}
\caption{\Planck\ SZ constraints (+BAO+BBN) on $\Lambda$CDM
  cosmological parameters in red. The black lines show the constraints
  upon substituting {\it HST} constraints on $H_0$ for the BAO
  constrainsts.  Contours are 68 and 95\% confidence levels.}
\label{fig:chainsnr7}
\end{figure*}

In the PSZ, the redshifts have been collected from different
observations and from the literature. Individual uncertainties in
redshift are thus spread between 0.001 and 0.1. Most of the clusters
in the cosmological sample have spectroscopic redshifts (184 out of
189) and we checked that the uncertainties in redshift are not at all
dominant in our error budget; they  are thus neglected. The cluster
without known redshift is incorporated by scaling the counts by a
factor 189/188, i.e., by assuming its redshift is drawn from the
distribution defined by the other 188 objects.

\subsection{Markov chain Monte Carlo}

In order to impose constraints on cosmological parameters from our
sample(s) given our modelled expected number counts, we modified {\tt
  CosmoMC} \citep{Lewis02} to include the likelihood described above.
We mainly study constraints on the spatially-flat $\Lambda$CDM model,
varying $\Omega_{\mathrm{m}}$, $\sigma_8$, $\Omega_{\mathrm{b}}$,
$H_0$, and $n_{\mathrm{s}}$, but also adding in the total neutrino
mass, $\sum m_{\nu}$, in Sect.~\ref{sec:discussion}.  In each of the
runs, the nuisance parameters ($Y_\ast$, $\alpha$, $\sigma_{\log Y}$)
follow Gaussian priors, with the characteristics detailed in
Table~\ref{table:scaling}, and are marginalized over.  The bias
  $(1-b)$ follows a flat prior in the range [0.7,1]. The redshift evolution of
the scaling, $\beta$, is fixed to its reference value unless stated
otherwise.

\subsection{External datasets \label{sec:external}}

When probing the six parameters of the $\Lambda$CDM model, we combine
the \Planck\ clusters with the Big Bang nucleosynthesis (BBN)
constraints from \citet{Steig08},
$\Omega_{\mathrm{b}}h^2=0.022\pm0.002$. We also use either the $H_0$
determination from {\it HST\/}  by \citet{Riess11}, $H_0=(73.8\pm2.4)$
kms$^{-1}$Mpc$^{-1}$, or baryon acoustic oscillation (BAO) data. In
the latter case we adopt the combined likelihood of \citet{wmap9} and
\citet{planck2013-p11}, which uses the radial BAO scales observed by
6dFGRS \citep{Beut11}, SDSS-DR7-rec and SDSS-DR9-rec
\citep{Pad12,And12}, and WiggleZ \citep{Blake12}.

\section{Constraints from \Planck\ clusters: $\Lambda$CDM \label{sec:constraints}} 

\subsection{Results for $\Omega_\mathrm{m}$ and $\sigma_8$}

\begin{figure}
\centering
\includegraphics[width=8.8cm]{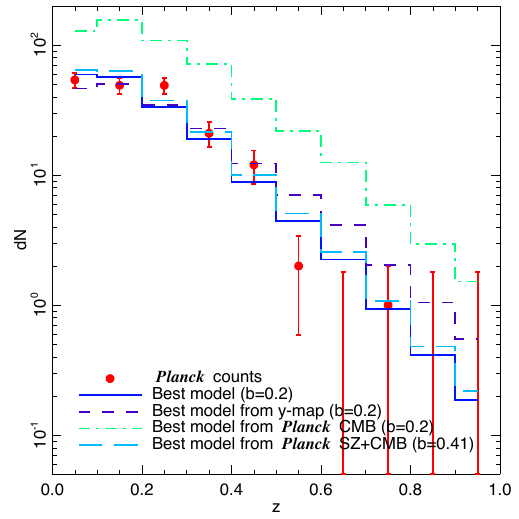} 
\caption{Distribution in redshift for the \Planck\ cosmological
  cluster sample. The observed number counts (red), are compared to
  our best-fit model prediction (blue). The dashed and dot-dashed
  histograms are the best-fit models from the \Planck\ SZ power
  spectrum and \Planck\ CMB power spectrum fits, respectively. The
cyan long dashed histogram is the best fit CMB+SZ when the bias is
  free (see Section~\ref{cmb}). The uncertainties on the observed
  counts, shown for illustration only, are the standard deviation
  based on the observed counts, except for empty bins where we show
  the inferred 84\% upper limit on the predicted counts assuming a
  Poissonian distribution. See Sect.~\ref{sec:discussion} for more
  discussion.}
\label{fig:dNdzsnr7}
\end{figure}

\begin{figure}
\centering
\includegraphics[width=8.8cm]{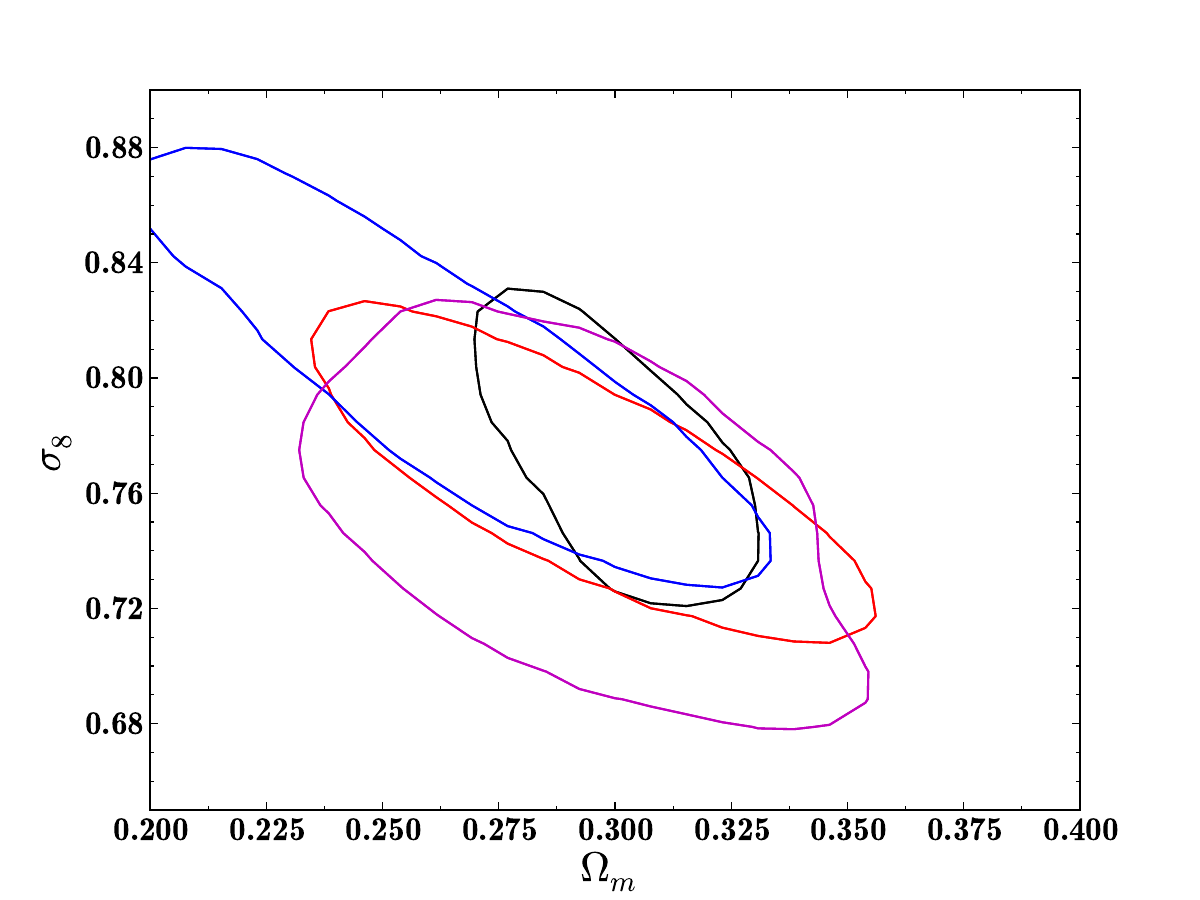} 
\caption{Comparison of the constraints using the mass functions of
  Watson et al.\ (black) and Tinker et al.\ (red), with $b$ fixed to 0.8.  When relaxing the constraints on the evolution of the
  scaling law with redshift (blue), the contours move along the
  degeneracy line. Allowing the bias
  to vary uniformly in the range $[0.7,1.0]$ enlarges the constraints
  perpendicular to the $\sigma_8$--$\Omega_\mathrm{m}$ degeneracy line
  due to the degeneracy of the number of clusters with the mass bias
  (purple). Contours are 95\% confidence levels here.}
\label{fig:diffmassfunc}
\end{figure}

Cluster counts in redshift for our \Planck\ cosmological sample are
not sensitive to all parameters of the $\Lambda$CDM model. We focus
first on ($\Omega_\mathrm{m}, \sigma_8$), assuming that $n_\mathrm{s}$
follows a Gaussian prior centred on the best-fit \Planck\ CMB
value\footnote{Table 2 of \citet{planck2013-p11}.}
($n_\mathrm{s}=0.9603\pm0.0073$).  We combine our SZ counts likelihood
with the BAO and BBN likelihoods discussed earlier. We also incorporate the
uncertainties on scaling parameters  in
Table~\ref{table:scaling}.   Allowing the bias to range uniformly over the interval
  $[0.7,1.0]$, we find the expected degeneracy between the two
  parameters,
  $\sigma_8(\Omega_\mathrm{m}/0.27)^{0.3}=0.764\pm0.025$,\footnote{We
    express it this way to ease comparison with other work.} with
  central values and uncertainties of
  $\Omega_\mathrm{m}=0.29\pm0.02$ and $\sigma_8=0.75\pm 0.03$
  (Table~\ref{tab:constraints} and Fig.~\ref{fig:chainsnr7}, red contours). The cluster counts
  as a function of redshift for the best-fit model are plotted in
  Fig.~\ref{fig:dNdzsnr7}.
When fixing the bias to $(1-b)=0.8$, the
  constraint on $\Omega_\mathrm{m}$ remains unchanged while the
  constraint on $\sigma_8$ becomes stronger:
  $\sigma_8(\Omega_\mathrm{m}/0.27)^{0.3}=0.78\pm0.01$ and
  $\sigma_8=0.77\pm 0.02$ (Fig.~\ref{fig:diffmassfunc}).

To investigate how robust our results are when changing our priors, we
repeat the analysis substituting the {\it HST\/} constraints on $H_0$
for the BAO results. Figure~\ref{fig:chainsnr7} (black contours) shows
that the main effect is to change the best-fit value of $H_0$, leaving
the $(\Omega_\mathrm{m}, \sigma_8)$ degeneracy almost unchanged.  In
the following robustness tests, we assume a fixed mass bias,
$(1-b)=0.8$, to better identify the effect of each of our assumptions.

\subsection{Robustness to the observational sample \label{sec:errorbudget}}

\begin{figure}
\centering
\includegraphics[width=8.8cm]{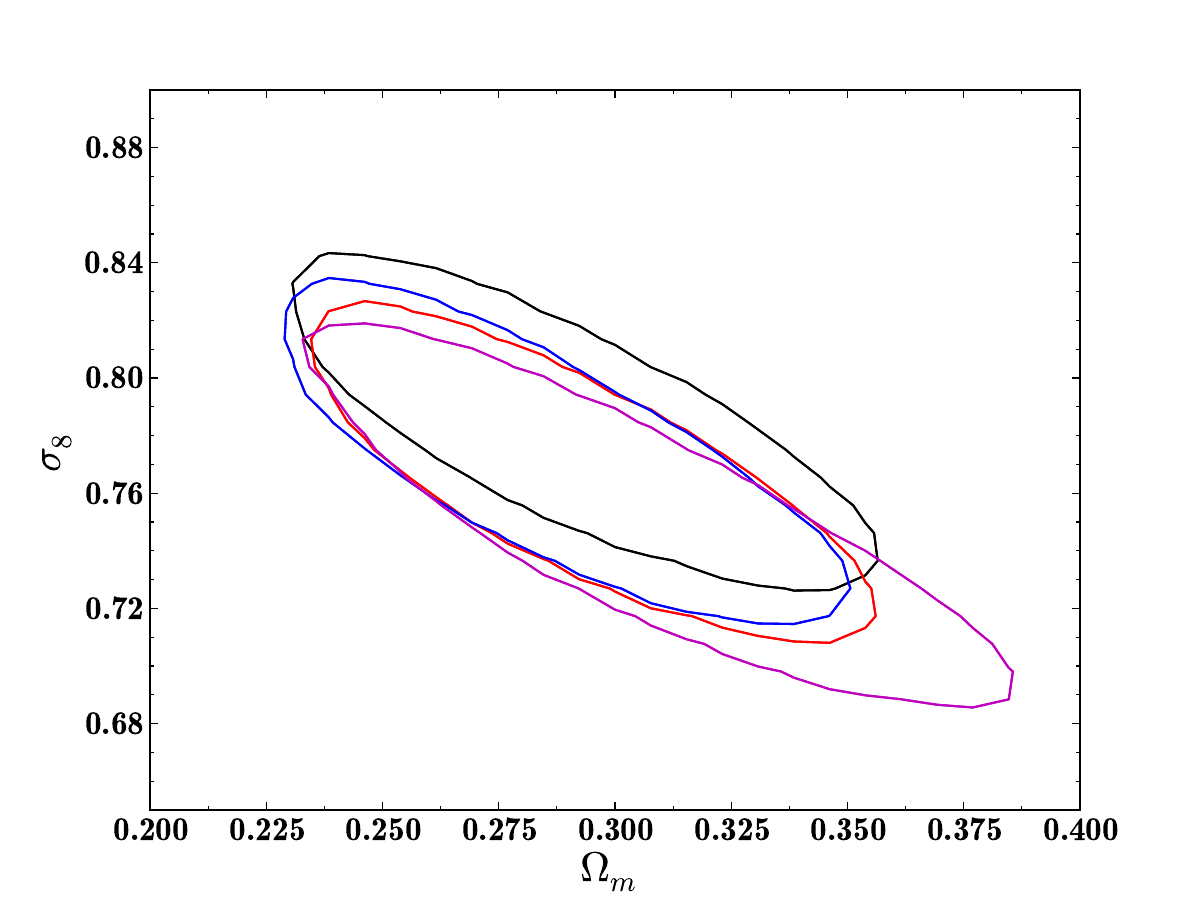}
\caption{95\% contours for different robustness tests: \texttt{MMF3}
  with S/N cut $>7$ in red; \texttt{MMF3} with S/N cut $>8$ in blue;
  and \texttt{MMF1} with S/N cut $>7$ in black; and \texttt{MMF3} with S/N
  cut at 7 but adopting the MC completeness in purple.}
\label{fig:diffsamples}
\end{figure}

To test the robustness of our results, we performed the same analysis
with different sub-samples drawn from our cosmological sample or from
the PSZ, as described in Sect.~\ref{sec:sampleselfunc}, following that
section's discussion of completeness.  Figure~\ref{fig:diffsamples}
shows the likelihood contours of the three samples (blue, \texttt{MMF3}
S/N $> 8$; red, \texttt{MMF3} S/N $> 7$; black, \texttt{MMF1} S/N $>
7$) in the ($\Omega_\mathrm{m}, \sigma_8$) plane. There is good
agreement between the three samples. Obviously the three samples are
not independent, as many clusters are common, but the noise estimates
for \texttt{MMF3} and \texttt{MMF1} are different leading to different
selection functions. Table~\ref{tab:constraints} summarizes the
best-fit values.

We perform the same analysis as on the baseline cosmological sample
(SZ+BAO+BBN), but employing a different computation of the completeness
function using the Monte Carlo method described in
Sect.~\ref{sec:sampleselfunc}. Figure~\ref{fig:diffsamples} shows the
change in the 2D likelihoods when this alternative approach is adopted.
The Monte Carlo estimation (in purple), being close to the analytic
one, gives constraints that are similar, but enlarge the contour along
the ($\Omega_\mathrm{m},\sigma_8$) degeneracy.

\subsection{Robustness to cluster modelling\label{sec:robust2}}

A key ingredient in the modelling of the number counts is the mass
function. Our main results adopt the Tinker et al.\ mass function as
the reference model. We compare these to results from the Watson et
al. \ mass function to evaluate the impact of uncertainty in
predictions for the abundance of the most massive and extreme
clusters.  Figure~\ref{fig:diffmassfunc} shows the 95\% contours when
adopting the different mass functions. The main effect is to change
the orientation of the degeneracy between $\Omega_\mathrm{m}$ and
$\sigma_8$, moving the best-fit values by less than $1\,\sigma$.

We also relax the assumption of standard evolution of the scalings
with redshift by allowing $\beta$ to vary with a Gaussian prior taken
from \citet{planck2011-5.2a}, $\beta = 0.66 \pm 0.5$. Once again, the
contours move along the $\sigma_8$--$\Omega_\mathrm{m}$ degeneracy
direction (shown in blue in Fig.~\ref{fig:diffmassfunc}).

As shown in Appendix~\ref{app:scaling}, the estimation of the mass
bias is not trivial and there is a large scatter amongst
simulations. The red and purple contours compare the different
constraints when fixing the mass bias to 0.8 and when allowing it to
vary uniformly in the range $[0.7,1.0]$ respectively. Modelling of the
cluster observable--mass relation is clearly the limiting factor in our
analysis.

\section{Discussion\label{sec:discussion}}

Our main result is the constraint in the
$(\Omega_\mathrm{m},\sigma_8)$ plane for the standard $\Lambda$CDM
model imposed by the SZ counts, which we have shown is robust to the
details of our modelling. We now compare this result first to
constraints from other cluster samples, and then to the constraints
from the \Planck\ analysis of the sky-map of the Compton $y$-parameter
\citep{planck2013-p05b} and of the primary CMB temperature
anisotropies \citep{planck2013-p11}.

\subsection{Comparison with other cluster constraints}

We restrict our comparison to some recent analyses exploiting a range
of observational techniques to obtain cluster samples and mass
calibrations.

\citet{Benson2011} used 18 galaxy clusters in the first
$178\,{\mathrm{deg}}^2$ of the SPT survey to find
$\sigma_{8}(\Omega_{\mathrm{m}}/0.25)^{0.3}=0.785\pm 0.037$ for a
spatially-flat model. They break the degeneracy between $\sigma_{8}$
and $\Omega_{\mathrm{m}}$ by incorporating primary CMB constraints,
deducing that $\sigma_8=0.795\pm 0.016$ and
$\Omega_{\mathrm{m}}=0.255\pm 0.016$. In addition, they find that the
dark energy equation of state is constrained to $w=-1.09\pm 0.36$,
using just their cluster sample along with the same {\it HST\/} and
BBN constraints used here. Subsequently, \citet{Reichardt2012}
reported a much larger cluster sample and used this to improve on the
statistical uncertainties on the cosmological parameters (see
Table~\ref{tab:constraints2}). \citet{Hass13} use a sample of 15 high
S/N clusters from ACT, in combination with primary CMB data, to find
$\sigma_8 = 0.786 \pm 0.013$ and $\Omega_{\mathrm{m}} = 0.250 \pm
0.012$ when assuming a scaling law derived from the universal pressure
profile.

\begin{figure}
\centering
\includegraphics[width=8.8cm]{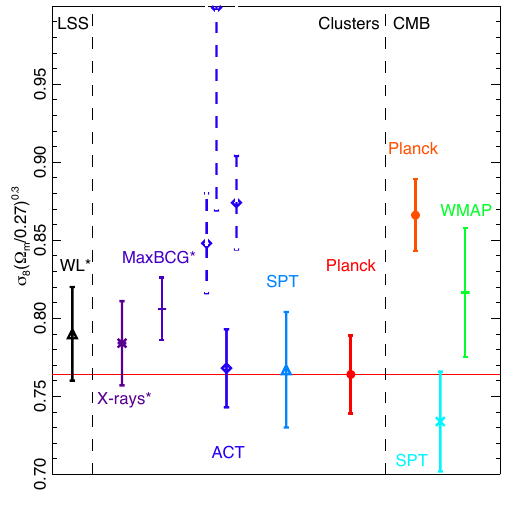}
\caption{Comparison of constraints (68\% confidence interval) on
  $\sigma_8 (\Omega_\mathrm{m}/0.27)^{0.3}$ from different experiments
  of large--scale structure (LSS), clusters, and CMB. The solid line ACT
  point assumes the same universal pressure profile as this
  work. Probes marked with an asterisk have an original power of
  $\Omega_\mathrm{m}$ different from 0.3. See text and 
  Table~\ref{tab:constraints2} for more details.}
\label{fig:omsigconstraints}
\end{figure}

\begin{table*}[tmb]                 % table* is a two-column table.  Drop the * for one column.
\begingroup
\newdimen\tblskip \tblskip=5pt
\caption{\footnotesize{Constraints from clusters on $\sigma_8(\Omega_\mathrm{m}/0.27)^{0.3}$.  } }
\label{tab:constraints2}                            % Label goes here.
\nointerlineskip
\vskip -3mm
\footnotesize
\setbox\tablebox=\vbox{
   \newdimen\digitwidth 
   \setbox0=\hbox{\rm 0} 
   \digitwidth=\wd0 
   \catcode`*=\active 
   \def*{\kern\digitwidth}
   \newdimen\signwidth 
   \setbox0=\hbox{+} 
   \signwidth=\wd0 
   \catcode`!=\active 
   \def!{\kern\signwidth}
\halign{ #\hfil\tabskip=2em & \hfil#\hfil & \hfil#\hfil & \hfil#\hfil & \hfil#\hfil & \hfil#\hfil \tabskip=0pt\cr
\noalign{\doubleline} \cr
Experiment                             & CPPP$^a$                          & MaxBCG$^b$                &         ACT$^c$         & SPT                 &  {\it Planck} SZ\cr 
\noalign{\vskip 3pt\hrule\vskip 5pt}
Reference                              &  Vikhlinin et al.                &    Rozo et al.            &  Hasselfield et al.     & Reichardt et al.   &     This work   \cr
 & (2009b) & (2010)  & (2013) & (2013) & \cr  
Number of clusters                     &           49+37                   &    $\sim$13000                  &               15        & 100                &           189   \cr 
Redshift range                         & $[0.025,0.25]$ and $[0.35,0.9]$       & $[0.1,0.3]$                &           $[0.2,1.5]$      & $[0.3,1.35]$         &   $[0.0,0.99]$    \cr 
Median mass ($10^{14} h^{-1} \msol$)     & 2.5                               & 1.5                      & 3.2                      & 3.3                & 6.0             \cr
Probe                                  &  $N(z,M)$                         &  $N(M)$             & $N(z,M)$                 & $N(z,Y_\mathrm{X})$         & $N(z)$               \cr 
S/N cut                                &      5                            &      ($N_{200}>11$)       &                 5        & 5                  & 7                \cr 
Scaling                                & $Y_{\mathrm{X}}$--$T_{\mathrm{X}}$, $M_{\mathrm{gas}}$  &   $N_{200}$--$M_{200}$       & several      &    $L_{\mathrm{X}}$--$M$, $Y_{\mathrm{X}}$        & $Y_{\mathrm{SZ}}$--$Y_{\mathrm{X}}$  \cr  
$\sigma_8(\Omega_\mathrm{m}/0.27)^{0.3}$  &      $0.784 \pm 0.027$     &      $0.806 \pm 0.033$  & $0.768 \pm 0.025$       &  $0.767 \pm 0.037$  &        $0.764 \pm 0.025$ \cr 
\noalign{\vskip 5pt\hrule\vskip 3pt}}}
%\endPlancktable                    % ends one-column \halign
\endPlancktablewide                 % ends two-column \halign
\tablenote a The degeneracy is  $\sigma_8(\Omega_\mathrm{m}/0.27)^{0.47}$.\par
\tablenote b The degeneracy is  $\sigma_8(\Omega_\mathrm{m}/0.27)^{0.41}$.\par
\tablenote c For ACT we choose the results assuming the scaling law derived from the universal pressure profile  in this table (constraints using other scaling relations are shown in Fig.~\ref{fig:omsigconstraints}).\par
\endgroup
\end{table*}                        % table* is a two-column table.  Drop the * for one column.

Strong constraints on cosmological parameters have been inferred from
X-ray and optical richness selected
samples. \citet{vik09} used a sample of 86 well-studied
X-ray clusters, split into low- and high-redshift bins, to conclude
that $\Omega_{\Lambda}>0$ with a significance about $5\,\sigma$ and
that $w=-1.14\pm 0.21$. \citet{2010ApJ...708..645R} used the
approximately $10^4$ clusters in the Sloan Digital Sky Survey (SDSS)
MaxBCG cluster sample, which are detected using a colour--magnitude
technique and characterized by optical richness. They found that
$\sigma_8(\Omega_{\mathrm{m}}/0.25)^{0.41}=0.832\pm 0.033$. The fact
that this uncertainty is similar to those quoted above for much
smaller cluster samples signifies, once again, that cluster cosmology
constraints are now limited by modelling, rather than statistical,
uncertainties.

Table~\ref{tab:constraints2} and Fig.~\ref{fig:omsigconstraints}
show some current constraints on the combination $\sigma_8
(\Omega_\mathrm{m}/0.27)^{0.3}$, which is the main degeneracy line in
cluster constraints.  This comparison is only meant to be
  indicative, as a more quantitative comparison would require full
  consideration of modelling details which is beyond the scope of this
  work. Cosmic shear \citep{kil13}, X-rays \citep{vik09}, and MaxBCG
\citep{2010ApJ...708..645R} each have a different slope in
$\Omega_\mathrm{m}$, being 0.6, 0.47, and 0.41, respectively (instead
of 0.3), as they are probing different redshift ranges. We have
rescaled when necessary the best value and errors to quote numbers
with a pivot $\Omega_\mathrm{m}=0.27$. \citet{Hass13} have derived
`cluster-only' constraints from ACT by adopting several different
scaling laws, shown in blue and dashed blue in
Fig.~\ref{fig:omsigconstraints}. The constraint assuming the universal
pressure profile is highlighted as the solid symbol and error bar. For
SPT we show the `cluster-only' constraints from
\citet{Reichardt2012}. For our own
analysis we show our baseline result for SZ+BAO+BBN with a prior on
$(1-b)$ distributed uniformly in $[0.7,1]$.
The figure thus demonstrates good
agreement amongst all cluster observations, whether in optical,
X-rays, or SZ. Table~\ref{tab:constraints2} compares the different
data and assumptions of the different cluster-related publications.

\subsection{Consistency with the \Planck\ y-map}

In a companion paper \citep{planck2013-p05b}, we performed an analysis
of the SZ angular power spectrum derived from the \Planck\ $y$-map
obtained with a dedicated component-separation technique. For the
first time, the power spectrum has been measured at intermediate
scales ($50 \le \ell \le 1000$). The same modelling as in
Sect.~\ref{sec:modelling} and \citet{Tab09,Tab10} has been used to
derive best-fit values of $\Omega_\mathrm{m}$ and $\sigma_8$, assuming
the universal pressure profile \citep{arn10}, a bias $1-b=0.8$, and
the best-fit values for other cosmological parameters from
\citet{planck2013-p11}.\footnote{For \Planck\ CMB we took the
  constraints from the \Planck+WP case, column 6 of Table 2 of
  \citet{planck2013-p11}. The baseline model includes massive
  neutrinos with $\sum m_{\nu}=0.06 \,$eV.} The best model obtained,
shown in Fig.~\ref{fig:dNdzsnr7} as the dashed line, demonstrates the
consistency between the PSZ number counts and the signal
observed in the $y$-map.

\subsection{Comparison with \Planck\ primary CMB constraints\label{cmb}}

We now compare the PSZ cluster constraints to those from the
analysis of the primary CMB temperature anisotropies given in
\citet{planck2013-p11} (see Footnote 6). In that analysis $\sigma_8$
is derived from the standard six $\Lambda$CDM parameters.

\begin{figure}
\centering
\includegraphics[width=8.8cm]{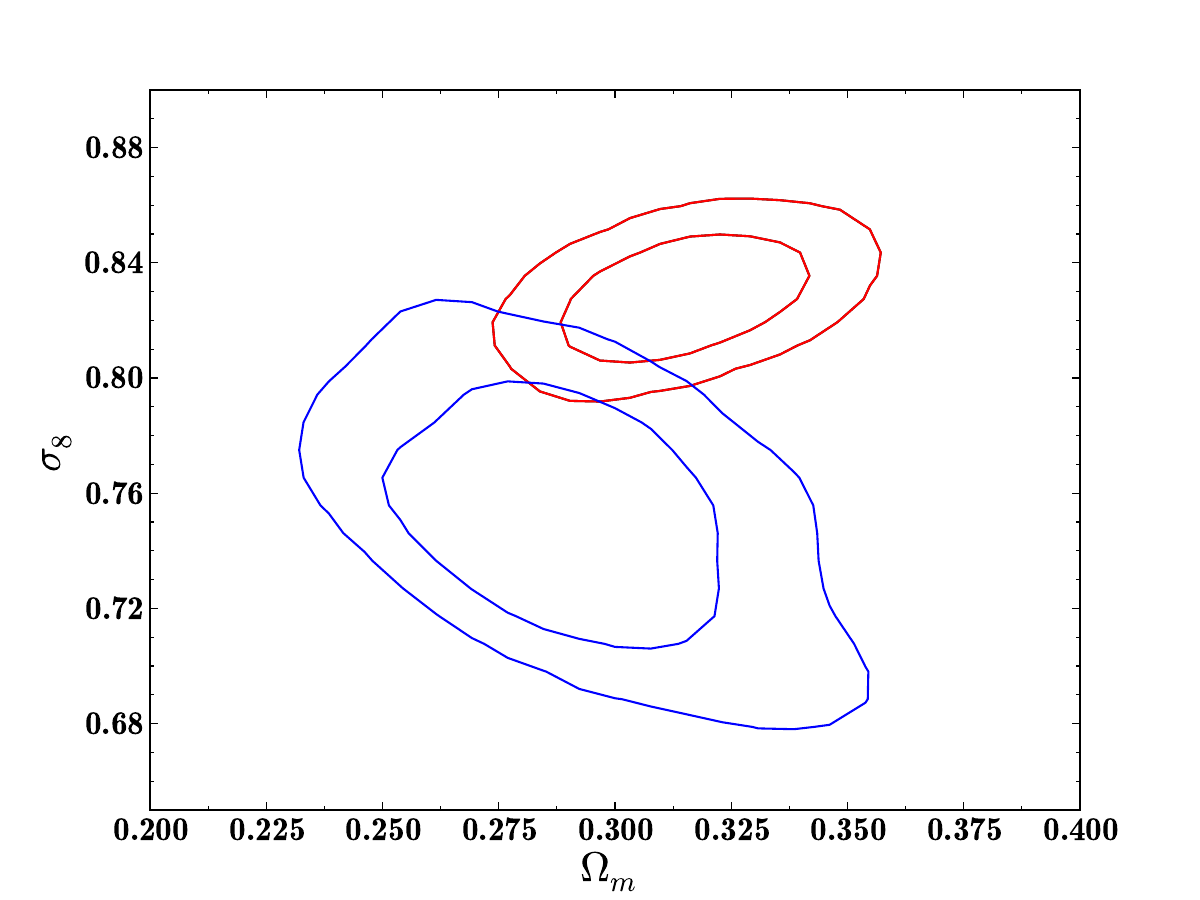}
\caption{ 2D $\Omega_\mathrm{m}$--$\sigma_8$ likelihood contours for the
  analysis with {\it Planck} CMB only (red); {\it Planck} SZ + BAO +
  BBN (blue) with $(1-b)$ in $[0.7,1]$. }
\label{fig:cont3}
\end{figure}

The \Planck\ primary CMB constraints, in the
$(\Omega_\mathrm{m},\sigma_8)$ plane, differ significantly from our
own, in particular through favouring a higher value of $\sigma_8$,
(see Fig.~\ref{fig:cont3}). For $(1-b)=0.8$, this leads to a factor of
$∼ 2$ larger number of predicted clusters than is actually observed
(see Fig.~\ref{fig:dNdzsnr7}). There is therefore some tension between
the results from the \Planck\ CMB analysis and the current cluster
analysis.  Figure~\ref{fig:omsigconstraints} illustrates this with a
comparison of three analyses of primary CMB data alone
\citep{planck2013-p11,Sto12,wmap9} and cluster constraints in terms of
$\sigma_8 (\Omega_\mathrm{m}/0.27)^{0.3}$.

It is possible that the tension results from a combination of some
residual systematics with a substantial statistical
fluctuation. Enough tests and comparisons have been  made on the
\Planck\ data sets that it is plausible that at least one discrepancy
at the two or three sigma level will arise by chance. Nevertheless, it is
worth considering the implications if the discrepancy is real.

As we have noted, the modelling of the cluster gas physics
  is the most important uncertainty in our analysis,  in
    particular through its influence on the mass bias $(1-b)$. While
    we have argued for a preferred value of \mbox{$(1-b) \simeq 0.8$}
    based on comparison of our $Y_{500}$--$M_{500}$ relation to those
    derived from a number of different numerical simulations, and we
    suggest a plausible range of $(1-b)$ from 0.7 to 1, a
    significantly lower value would substantially alleviate the
    tension between CMB and SZ constraints. We have undertaken a joint
    analysis using the CMB likelihood presented in
    \citet{planck2013-p08} and the cluster likelihood presented in the
    present paper, sampling $(1-b)$ in the range $[0.1,1.5]$. This
    results in a `measurement' of $(1-b)=0.59 \pm 0.05$.  We show in
    Fig~\ref{fig:dNdzsnr7} the SZ cluster counts predicted by the \Planck's
    best-fit primary CMB model for $(1-b)=0.59$. Clearly, this
    substantial reduction in $(1-b)$ is enough to reconcile our observed
    SZ cluster counts with \Planck's best fit primary CMB parameters.
  
  {Such a large bias is difficult to reconcile with numerical
      simulations, and cluster masses estimated from X-rays and from
      weak lensing do not typically show such large offsets (see
      Appendix~\ref{app:scaling}). Systematic discrepancies in the
      relevant scaling relations have, however, been identified and
      studied in stacking analyses of X-ray, SZ, and lensing data for
      the very large MaxBCG cluster sample, e.g.,
      \citet{planck2011-5.2c}, \citet{2012ApJ...757....1B},
      \citet{2012PhRvD..85b3005D}, \citet{2012ApJ...760...67R}, and
      \citet{2013ApJ...767...38S}, suggesting that the issue is not
      yet fully settled from an observational point of view. The
      uncertainty reflects the inherent biases of the different
      mass estimates. Systematic effects arising from
      instrument calibration constitute a further source of
      uncertainty --- in X-ray mass determinations, temperature
      estimates represent the main source of systematic uncertainty in
      mass, as the mass scales roughly with $T^{3/2}$. Other biases in
      the determination of mass--observable scaling relations come
      from the object selection process itself
      \citep[e.g.,][]{man10,all11,ang12}. This may be less of a
      concern for SZ selected samples because of the expected small
      scatter between the measured quantity and the mass. An improbable
      conspiracy of all sources of bias seems required to lead to a
      sufficiently-low effective value of $(1-b)$ to reconcile the SZ and
      CMB constraints. This possibility needs to be carefully examined
      with probes based on a variety of physical quantities and
      derived from a wide range of types of observation, including
      masses, baryon and gas fractions, etc. }

A different mass function may also help reduce the tension.  Mass
functions are calibrated against numerical simulations that may still
suffer from volume effects for the largest haloes, as shown in the
difference between the \citet{2008ApJ...688..709T} and \citet{wat12}
mass functions.  This does not seem sufficient, however, given the
results presented in Fig.~\ref{fig:diffmassfunc}.

One might instead ask whether the \planck\ data analysis could
  somehow have missed a non-negligible fraction of the total number of
  clusters currently predicted to have $\mathrm{S/N} > 7$, resulting
  in a lower observed number count distribution. This is linked to a
  possible underestimate of the true dispersion about the
  $\Yv$--$Y_{\mathrm{X}}$ relation at a given $\Mv$. It would be
  necessary for \planck\ to have missed $\sim$40 percent of the
  clusters with  predicted SZ $\mathrm{S/N} > 7$ in order for
  the SZ and CMB number count curves in Fig.~\ref{fig:dNdzsnr7} to be
  in agreement. Increasing the dispersion about the $\Yv$--$\Mv$
  relation and allowing it to correlate strongly with the scatter in
  X-ray properties (in particular, $Y_{\mathrm{X}}$) would raise the
  possibility that our calibration procedures (which are based on
  X-ray and SZ selected clusters assuming the scatter in $Y$ and
  $Y_{\mathrm{x}}$ at fixed $\Mv$ to be small and to be uncorrelated
  with cluster dynamical state) might produce a relation which is
  biased high. A sufficiently-large effect seems, however, to require
  a level of scatter and a degree of correlation with cluster
  structure which are inconsistent with the predictions of current
  hydrodynamical simulations (see the discussion in
  Appendix~\ref{app:scaling}).

\begin{figure}
\centering
\includegraphics[width=8.8cm]{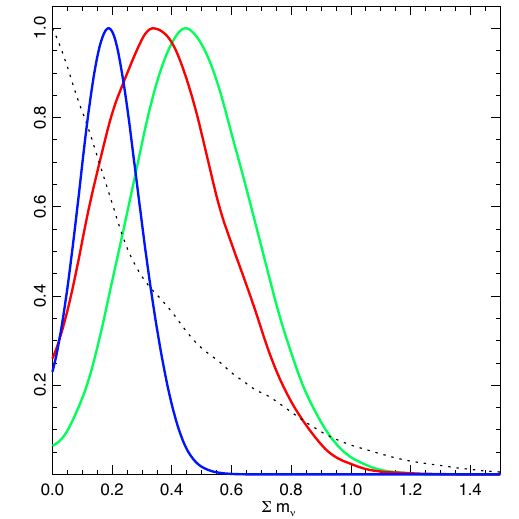}
\caption{Marginalized posterior distribution for $\sum m_{\nu}$ from: {\it Planck} CMB data alone (black dotted line); {\it Planck} CMB + SZ with $1-b$ in $[0.7,1]$ (red); {\it Planck} CMB +
  SZ + BAO with $1-b$ in $[0.7,1]$ (blue); and {\it Planck} CMB + SZ with
  $1-b=0.8$ (green).}
\label{fig:nu}
\end{figure}

Alternatively, the discrepancy may reflect a need to extend the
minimal $\Lambda$CDM model in which the $\sigma_8$ constraints are derived
from the primary CMB analysis. Any extension would need to modify the
power spectrum on the scales probed by clusters, while leaving the
scales probed by primary CMB observations unaffected. The inclusion of
neutrino masses, quantified by their sum over all families, $\sum
m_{\nu}$, can achieve this (see \citealt{2011MNRAS.418..346M} and
\citealt{bur13} for reviews of how cosmological observations can be
affected by the inclusion of neutrino masses).  The SPT collaboration
\citep{hou14} recently considered such a possibility to mitigate their
tension with WMAP-7 primary CMB data.  There is an upper limit
of $\sum m_{\nu}<0.93\,$eV from the \Planck\ primary CMB data
alone.\footnote{ \citet{planck2013-p11}, Table 10, column 3.} If we
combine the \Planck\ CMB (\Planck+WP) likelihood and the cluster count
data using a fixed value $(1-b)=0.8$,  then we find a
  $2.8\,\sigma$ preference for the inclusion of neutrino masses with
  $\sum m_{\nu}=(0.53 \pm 0.19)\,$eV, as shown in
Fig.~\ref{fig:nu}. If, on the other hand, we adopt a more conservative
point of view and allow $(1-b)$ to vary between 0.7 and 1.0, this
preference drops to  $1.9 \sigma$ with $\sum m_{\nu}=(0.40 \pm
  0.21)\,$eV. Adding BAO data to the compilation lowers the value of
the required mass but increases the significance, yielding  $\sum
  m_{\nu}=(0.20 \pm 0.09)\,$eV, due to a breaking of the degeneracy
between $H_0$ and $\sum m_\nu$.

As these results depend on the value and allowed range of $(1-b)$,
better understanding of the scaling relation is the key to further
investigation.  This provides strong motivation for further study of
the relationship between $Y$ and $M$.  Over the past few years we
  have moved into an era where \textit{systematic} uncertainties
  dominate to an increasing extent over \textit{statistical}
  uncertainties. Observed mass estimates using different methods are
  of improving quality; for instance, X-ray mass proxies can
  be measured to better than 10\% precision
  \citep[e.g.,][]{vik06, arn07}. In this context, systematic
  calibration uncertainties are playing an increasingly important role
  when using the cluster population to constrain cosmology.

\section{Summary}

We have used a sample of nearly 200 clusters from the PSZ, along with
the corresponding selection function, to place strong constraints in
the $(\Omega_\mathrm{m},\sigma_8)$ plane.  We have carried out a
series of tests to verify the robustness of our constraints, varying
the observed sample choice, the estimation method for the selection
function, and the theoretical methodology, and have found that our
results are not altered significantly by those changes.

The relation between the mass and the integrated SZ signal plays a
major role in the computation of the expected number
counts. Uncertainties in cosmological constraints from clusters are no
longer dominated by small number statistics, but by the gas physics
and sample selection biases. Uncertainties in the $Y$--$M$ relation
include contributions from X-ray instrument calibration, X-ray
temperature measurements, inhomogeneities in cluster density and
temperature profiles, and selection effects. Considering several
ingredients of the gas physics of clusters, numerical simulations
predict scaling relations with 30\% scatter in amplitude (at a
fiducial mass of $6 \times 10^{14} \msol$), and suggest a mass bias
between the true mass and the estimated mass of $(1-b) =
0.8^{+0.2}_{-0.1}$. Adopting the central value we found constraints on
$\Omega_\mathrm{m}$ and $\sigma_8$ that are in good agreement with
previous measurements using clusters of galaxies.

Comparing our results with \Planck\ primary CMB constraints within the
$\Lambda$CDM cosmology reveals some tension. This can be alleviated
by permitting a large mass bias ($1-b \simeq 0.60$), which is however
significantly larger than expected. Alternatively, the tension may
indicate a need for  an extension of the base $\Lambda$CDM model that modifies its
power spectrum shape. For example the inclusion of non-zero neutrino
masses helps in reconciling the primary CMB and cluster constraints, a
fit to \Planck\ CMB + SZ + BAO yielding $\sum m_{\nu}=(0.20 \pm
0.09)\,$eV.

Cosmological parameter determination using clusters is currently
limited by the knowledge of the observable--mass relations. In the
future our goal is to increase the number of dedicated follow-up
programmes to obtain better estimates of the mass proxy and redshift
for most of the S/N $>5$ \Planck\ clusters. This will allow for improved
determination of the scaling laws and the mass bias, increase the
number of clusters that can be used, and allow us to investigate an
extended cosmological parameter space.

\begin{acknowledgements}
The development of \Planck\ has been supported by: ESA; CNES and
CNRS/INSU-IN2P3-INP (France); ASI, CNR, and INAF (Italy); NASA and DoE
(USA); STFC and UKSA (UK); CSIC, MICINN, JA and RES (Spain); Tekes,
AoF and CSC (Finland); DLR and MPG (Germany); CSA (Canada); DTU Space
(Denmark); SER/SSO (Switzerland); RCN (Norway); SFI (Ireland);
FCT/MCTES (Portugal); and PRACE (EU). A description of the
Planck Collaboration and a list of its members, including the
technical or scientific activities in which they have been involved,
can be found at
\url{http://www.sciops.esa.int/index.php?project=planck&page=Planck_Collaboration}
\end{acknowledgements}

\bibliographystyle{aa}

\bibliography{Planck_bib,sz_cosmo_counts_bib}

\appendix

\section{Calibration of the \YM\ relation \label{app:scaling}}

A cluster catalogue is a list of positions and measurements of
observable physical quantities. Its scientific utility depends largely
on our ability to link the observed quantities to the underlying mass,
in other words, to define an observable proxy for the
mass. \planck\ detects clusters through the SZ effect. This effect is
currently the subject of much study in the cluster community, chiefly
because numerical simulations indicate that the spherically-integrated
SZ measurement is correlated particularly tightly with the underlying
mass. In other words, this measurement potentially represents a
particularly valuable mass proxy.

To establish a mass proxy, one obviously needs an accurate measurement
both of the total mass and of the observable quantity in
question. However, even with highly accurate measurements, the
correlation between the observable quantity and the mass is
susceptible to {\it bias} and {\it dispersion}, and both of these
effects need to be taken into account when using cluster catalogues
for scientific applications.

The aim of this Appendix is to define a baseline relation between the
measured SZ flux, $\YSZ$, and the total mass $\Mv$. The latter quantity
is not directly measurable. On an individual cluster basis, it can be
inferred from dynamical analysis of galaxies, from X-ray analysis
assuming hydrostatic equilibrium (HE), or from gravitational lensing
observations. However, it is important to note that {\it all} observed
mass estimates include inherent biases. For instance, numerical
simulations suggest that HE mass measurements are likely to
underestimate the true mass by $10$--$15$ percent due to neglect of
bulk motions and turbulence in the intra-cluster medium 
\citep[ICM, e.g.,][]{nag07,pif08,mene10}, an effect that is commonly
referred to in the literature as the `hydrostatic mass
bias'. Similarly, simulations indicate that weak lensing mass
measurements may underestimate the mass by 5 to 10 percent, owing to projection
effects or the use of inappropriate mass models
\citep[e.g.,][]{bec11}. Instrument calibration systematic effects
constitute a further source of error. For X-ray mass determinations,
temperature estimates represent the main source of systematic
uncertainty, as the mass at a given density contrast scales roughly
with $T^{3/2}$. Other biases in the determination of mass--observable
scaling relations come from the object selection process itself
\citep[e.g.,][]{all11,ang12}. A classic example is the Malmquist bias,
where bright objects near the flux limit are preferentially
detected. This effect is amplified by Eddington bias, the mass
function dictating that many more low-mass objects are detected
compared to high-mass objects. Both of these biases depend critically
on the distribution of objects in mass and redshift, and on the
dispersion in the relation between the mass and the observable used
for sample selection. This is less of a concern for SZ selected
samples than for X-ray selected samples, the SZ signal having much
less scatter at a given mass than the X-ray luminosity. However for
precise studies it should still be taken into account.

On the theoretical side, numerous $\Yv$--$\Mv$ relations have been
derived from simulated data, as discussed below. The obvious advantage
of using simulated data is that the relation between the SZ signal and
the true mass can be obtained, because the `real' value of all
physical quantities can be measured. The disadvantage is that the
`real' values of measurable physical quantities depend strongly on the
phenomenological models used to describe the different
non-gravitational processes at work in the ICM.

Nevertheless, the magnitude of the bias between observed and true
quantities can only be assessed by comparing multi-wavelength
observations of a well-controlled cluster sample to numerical
simulations. Thus, ideally, we would have full follow-up of a complete
\planck\ cluster sample. For large samples, however, full follow-up is
costly and time consuming. This has led to the widespread use of mass
estimates obtained from mass--proxy relations. These relations are
generally calibrated from individual deep observations of a subset of
the sample in question \citep[e.g.,][]{vik09b}, or from deep
observations of objects from an external dataset \citep[e.g., the use of
the \rexcess\ relations in][]{planck2011-5.2b}.

For the present paper, we will rely on mass estimates from a
mass--proxy relation.  In this context, the \MYX\ relation is clearly
the best to use. $\YX$, proposed by \citet{kra06}, is defined as the
product of $\Mgv$, the gas mass within $\Rv$, and $\TX$, the
spectroscopic temperature measured in the $[0.15$--$0.75]~\Rv$
aperture. In the simulations performed by \citet{kra06}, $\YX$ was
extremely tightly correlated with the true cluster mass, with a
logarithmic dispersion of only $8$ percent. Observations using masses
derived from X-ray hydrostatic analysis indicate that $\YX$ does
indeed appear to have a low dispersion \citep{arn07,vik09b}.
Furthermore, the local \MYX\ relation for X-ray selected relaxed
clusters has been calibrated to high statistical precision
\citep{arn10, vik09b}, with excellent agreement achieved between
various observations \citep[see e.g.,][]{arn07}. Since
  simulations suggest that the \YM\ relation is independent of
  dynamical state, calibrating the \YM\ relation via a low-scatter
mass proxy, itself calibrated on clusters for which the HE bias is
expected to be minimal, is a better approach than using HE mass
estimates for the full sample, since the latter can be highly
biased for very unrelaxed objects.

%%%%%%%%%%
% Y-M relation
%

\begin{table*}[tmb]                 % table* is a two-column table.  Drop the * for one column.
\begingroup
\newdimen\tblskip \tblskip=5pt
\caption{\footnotesize{ Parameters for the \YM\ relation, expressed as
    $E^{-2/3}(z)\left[D_\mathrm{A}^2
      \YSZ/10^{-4}{\mathrm{Mpc^2}}\right]=
    10^A \left[\Mv/6\times 10^{14} \msol \right]^\alpha$.  } }                          % Caption goes here.
\label{tab:ym}                            % Label goes here.
\nointerlineskip
\vskip -3mm
\footnotesize
\setbox\tablebox=\vbox{
   \newdimen\digitwidth 
   \setbox0=\hbox{\rm 0} 
   \digitwidth=\wd0 
   \catcode`*=\active 
   \def*{\kern\digitwidth}
   \newdimen\signwidth 
   \setbox0=\hbox{+} 
   \signwidth=\wd0 
   \catcode`!=\active 
   \def!{\kern\signwidth}
\halign{ #\hfil\tabskip=2em & \hfil#\hfil & \hfil#\hfil & \hfil#\hfil & \hfil#\hfil & \hfil#\hfil & \hfil#\hfil & \hfil#\hfil & \hfil#\hfil  \tabskip=0pt\cr
\noalign{\doubleline} \cr
Sample & $N_{\mathrm{ c}}$ &  MB & Mass &  {$A$} &  {$\alpha$}&
$ [\sigma_{\mathrm{log{Y | M}}}] $ int &$ [\sigma_{\mathrm{log{Y | M}}}] $ raw &  Section \cr 
\noalign{\vskip 3pt\hrule\vskip 5pt}
XMM-ESZ PEPXI &62& N & $\Mvy$			& $-0.19*\pm0.01*$ &
$1.74\pm0.08$&  $0.10*\pm0.01*$& ... & A.2.1 \cr 
\noalign{\vskip 3pt}
Cosmo sample   &  71 & N & $\Mvy$                           &
$-0.175\pm0.011$ & $1.77 \pm0.06$&  $0.065\pm0.010$&
$0.080\pm0.009$ & A.2.1 \cr 
\noalign{\vskip 3pt}
\bf
Cosmo sample   &  \bf 71 & \bf Y & $\mathbf{\mathit{ \Mvy}}$
&${\mathbf {-0.186\pm0.011}}$ &${ \mathbf {1.79 \pm0.06}} $&${
  \mathbf {0.063\pm0.011}}$&  ${\mathbf {0.079\pm0.009}}$ &  {\bf A.2.2} \cr
\noalign{\vskip 3pt}
XMM-ESZ  & 62 & Y & $\Mvy$                                &
$-0.19*\pm0.01*$ & $1.75\pm0.07$&  $0.065\pm0.011$& $0.079\pm0.009$ &  A.2.3 \cr 
\noalign{\vskip 3pt}
S/N $>7$  & $78$ & Y & $\Mvy$ &$-0.18*\pm0.01*$ & $1.72\pm0.06$&
$0.063\pm0.010$& $0.078\pm0.008$ &  A.2.3 \cr
\noalign{\vskip 3pt}
Cosmo sub-sample A  & 10  & Y & $\Mvh$& $-0.15*\pm0.04*$ &
$1.6*\pm0.3*$&  ...& $0.08*\pm0.02*$ &  A.3.2\cr 
\noalign{\vskip 3pt}
Cosmo sub-sample B   & 58 & Y & $\Mvh$& $-0.19*\pm0.03*$ & $1.7*\pm0.2*$&
$0.25*\pm0.06*$& $0.27*\pm0.06*$ &  A.3.2 \cr  
\noalign{\vskip 5pt\hrule\vskip 3pt}}}
%\endPlancktable                    % ends one-column \halign
\endPlancktablewide                 % ends two-column \halign
%\tablenote a Footnote a.\par
%\tablenote b Footnote b.\par
\endgroup
\tablefoot{Column
    1, considered sample; column 2, number of clusters in the sample;
    column 3, Malmquist bias correction; if this column contains Y, a
    mean correction for Malmquist bias has been applied to each point
    before fitting; column 4, mass definition; columns 5 and 6, slope
    and normalization of the relation; columns 7 and 8, intrinsic and
    raw orthogonal scatter around the best-fit relation at a given mass; column 9, Section in which sub-sample is discussed.  The Cosmo sample highlighted in bold represents the baseline relation (see text for details).  }
\end{table*}                        % table* is a two-column table.  Drop the * for one column.

%%%%%%%%%

We approach the determination of the \YM\ relation in two steps. We
first calibrate the $\YSZ$--proxy relation. This is combined with the
X-ray calibrated relation, between the proxy and $\Mv$, to define an
observation-based \YM\ relation. In the second step, we assess
possible biases in the relation by directly comparing the
observation-based relation with that from simulations. This approach,
rather than directly assessing the HE mass bias, allows us to avoid
complications linked to the strong dependence of the HE bias on
cluster dynamical state, and thus on the cluster sample (real or
simulated).  The final output from this procedure is a relation
between $\Yv$ and $\Mv$, with a full accounting of the different
statistical and systematic uncertainties that go into its derivation,
including bias.

In the following, all relations are fit with a power law in log-space
using the orthogonal BCES method \citep{akr96}, which takes into account the
uncertainties in both variables and the intrinsic scatter. All
dispersions are given in log$_{10}$.

\subsection{Baseline mass--proxy relation}

As a baseline, we use the relation between $Y_{\mathrm{ X}}$ and the
X-ray hydrostatic mass $\Mv^{\mathrm{ HE}}$ established for 20 local
{\it relaxed} clusters by \citet{arn10}:
 \begin{eqnarray}
\label{eq:yx_m}
E^{-2/3}(z)\left[\frac{\YX}{2\times10^{14} \msol \,\keV}\right] \hspace*{3cm}\\
 \hspace*{2cm}           =  10^{0.376\pm0.018}
            \left[\frac{\Mvh}{6\times10^{14} \msol}\right]^{1.78\pm0.06},
            \nonumber 
\end{eqnarray}
assuming standard evolution, and where the uncertainties are
statistical only.  For easier comparison with the \YM\ relation given
below, the normalization for $\YX$ expressed in $10^{-4}\,{\mathrm{
    Mpc}}^2$ is $ 10^{-0.171\pm0.018}$.  The HE mass is expected to be
a biased estimator of the true mass,
\begin{equation}
\Mvh = (1-b)\,\Mv \,,
\label{eq:mb} 
\end{equation}
where all of the possible observational biases discussed above
(departure from HE, absolute instrument calibration, temperature
inhomogeneities, residual selection bias) have been subsumed
into the bias factor $(1-b)$. The form of the \mbox{\YXM}\ relation is
thus
\begin{equation}
E^{-2/3}(z)\YX=  10^{A\pm\sigma_{\mathrm{A}}} \,
\left[(1-b)\,\Mv\right]^{\alpha\pm \sigma_{\mathrm{\alpha}}} \,,
\label{eq:yx_mg}
\end{equation}
where $\sigma_{\mathrm{ A}}$ and $\sigma_{\mathrm{ \alpha}}$ are the
statistical uncertainties on the normalization and slope and $b$ is
the bias between the true mass and the observed mass used to calibrate
the relation. The bias is a poorly-known stochastic variable with
substantial variation expected between clusters. In our case, $b$
represents the {\it mean} bias between the observed mass and the true
mass.

The mass proxy $\Mvy$  is defined from the best-fit \YXMh\ relation
\begin{equation}
E^{-2/3}(z)\YX =  10^{A} \, \left[\Mvy\right]^{\alpha}\,.
\label{eq:myx}
\end{equation}
For any cluster, $\Mvy$, together with the corresponding $\YX$ and
$\Rvy$, can be estimated iteratively about this relation from the
observed temperature and gas mass profile, as described in
\citet{kra06}.  The calibration of the \YXM\ relation is equivalent to
a calibration of the $\Mvy$--$\Mv$ relation, which relates the mass
proxy, $\Mvy$, to the mass via
\begin{equation}
\Mvy  =  10^{\pm \sigma_{\mathrm{ A}}/\alpha}\,
\left[(1-b)\,\Mv\right]^{1\pm \sigma_{\mathrm{ \alpha}}/\alpha}\,. 
\label{eq:myx_m}
\end{equation}
In addition to the bias factor, there are statistical uncertainties on
the slope and normalization of the relation, as well as intrinsic
scatter around the relation, linked to the corresponding statistical
uncertainties and scatter of the \YXMh\ relation.

\subsection{Relation between $\YSZ$ and $\Mvy$}

\subsubsection{Best-fit relation}

We first investigate the relationship between $\YSZ$ and $\Mvy$, the
mass estimated iteratively from Eq.~\ref{eq:myx}, with parameters
given by the best-fit \citet{arn10} relation (Eq.~\ref{eq:yx_m}). Full
X-ray follow-up of the \planck\ SZ cosmological cluster sample is not
yet available. Our baseline sample is thus a subset of 71 detections
from the \planck\ cosmological cluster sample, detected at S/N $> 7$,
for which good quality \xmm\ observations are available. The sample
consists of data from our previous archival study of the
\Planck\ Early SZ (ESZ) clusters \citep{planck2011-5.2b}, of
\Planck-detected LoCuSS clusters presented by \citet{planck2012-III},
and from the \xmm\ validation programme
\citep[][]{planck2011-5.1b,planck2012-I,planck2012-IV}. The
corresponding sub-samples include 58, 4, and 9 clusters,
respectively.  The X-ray data were re-analysed in order to have a homogeneous data set; measurement differences are negligible with respect to previously-published values. Uncertainties on $\YX$, $\Rvy$, and $\Mvy$ include those
due to statistical errors on the X-ray temperature and the gas mass
profile. 

The SZ signal is estimated within a sphere of radius $\Rvy$ centred on
the position of the X-ray peak, as detailed in e.g.,
\citet[][]{planck2011-5.2b}. The re-extraction procedure uses matched
multi-filters (MMF) and assumes that the ICM pressure follows the
universal profile shape derived by \citet{arn10} from the combination
of the \rexcess\ sample with simulations.  The extraction is
undertaken on the 15.5 month \planck\ survey data set, and so
statistical precision on the SZ signal is improved with respect to
previously-published values.  The uncertainty on $\Yv$ includes
statistical uncertainties on the SZ signal derived from the MMF, plus
the statistical uncertainty on the aperture $\Rvy$. The latter
uncertainty is negligible compared to the statistical error on the SZ
signal.  The resulting relation for these 71 clusters from the
cosmological sample is
\begin{eqnarray}
\label{eq:y_myx}
E^{-2/3}(z)\left[\frac{D_{\mathrm{
          A}}^2\,\Yv}{\mathrm{10^{-4}\,Mpc^2}}\right] \hspace*{3cm} \\
\hspace*{2cm} = 10^{-0.175\pm0.011}
  \left[\frac{\Mv^{\YX}}{6\times10^{14}\msol}\right]^{1.77\pm0.06}. \nonumber
\end{eqnarray}
This agrees within $1\sigma$ with the results from the sample of 62
clusters from the ESZ sample with archival \xmm\ data published in
\citet{planck2011-5.2b}.  The slope and normalization are determined
at slightly higher precision, due to the better quality SZ data. The
derived intrinsic scatter (Table~\ref{tab:ym}) is significantly
smaller. This is a consequence of: a more robust treatment of
statistical uncertainties; propagation of gas mass profile
uncertainties in the $\YX$ error budget; and, to a lesser extent, the
propagation of $\Rvy$ uncertainties to $\YSZ$ estimates.

\subsubsection{Effects of Malmquist bias}

\begin{figure}
\includegraphics[width=8.8cm]{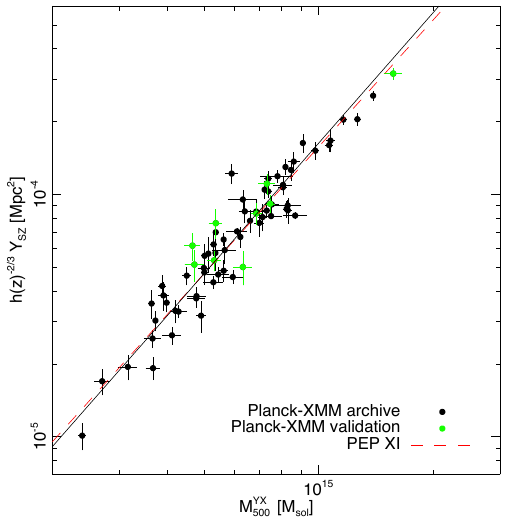}
\hfill
\caption{\label{fig:malrel} Best scaling relation between $\Yv$ and
  $\Mv$, and the  data points utilized after correction of the
  Malmquist bias}
\end{figure}

The fitted parameters are potentially subject to selection effects
such as Malmquist bias, owing to part of the sample lying close to the
selection cut. For the present sample, we use an approach adapted from
that described in \citet{vik09b} and \citet{pra09}, where each data
point is rescaled by the mean bias for its flux, and the relation
refitted using the rescaled points. The method is described in more
detail in Paper 1. For the baseline cosmological sample of 71 systems,
the bias-corrected $\Yv$--$\Mvy$ relation is
\begin{eqnarray}
\label{eqn:y_myxmb}  
E^{-2/3}(z)\left[\frac{D_{\mathrm{
          A}}^2\,\Yv}{\mathrm{ 10^{-4}\,Mpc^2}}\right]  \hspace*{3cm} \\
\hspace*{2cm} = 10^{-0.19\pm0.01}
  \left[\frac{\Mv^{\YX}}{6\times10^{14}\msol}\right]^{1.79\pm0.06}
  \,. \nonumber
\end{eqnarray}
The best-fit relation, together with Malmquist bias corrected data points,
is plotted in Fig.~\ref{fig:malrel}.

The correction decreases the effective $\YSZ$ values at a given mass,
an effect larger for clusters closer to the S/N threshold. The net
effect is small, a roughly $1\,\sigma$ decrease of the normalization and
a slight steepening of the power-law slope (Table~\ref{tab:ym}).

\subsubsection{Stability of slope and normalization}

The slope and normalization of this relation are robust
to the choice of sample (Table~\ref{tab:ym}). We compared our results to those obtained from:

  \begin{itemize}
  \item An extended sample of 78 clusters with $S/N > 7$ (71 in common
    with the baseline sample). This is built from all objects falling
    {in the 84\% sky mask used to define the SZ catalogue
      \citet{planck2013-p05a}}, and for which \xmm\ data have been
    published by the Planck Collaboration
    \citep{planck2011-5.1b,planck2011-5.2b,planck2012-I,planck2012-III,planck2012-IV}.

  \item The original 62 clusters from the ESZ sample published in
    \citet{planck2011-5.2b}, with updated SZ signal measurements
    obtained from 15.5 month \planck\ data (62 in common with the
    baseline sample). These objects are all known from X-ray surveys
    and all lie at $z < 0.5$. We use them to test fit robustness to
    the inclusion of non-X-ray selected, higher-redshift systems.
\end{itemize}

 As indicated in Table~\ref{tab:ym}, there is agreement within
  $1\,\sigma$ between the various samples.  The results are also in
  agreement with the relation obtained from a simple combination of
  the $\Yv$--$\YX$ relation (discussed in Paper 1) and the
  $\YX$--$\Mv^{\mathrm{ HE}}$ relation (Eq.~\ref{eq:yx_m} above).

\subsection{The observation-based  $\YSZ$--$\Mv$  relation}

\subsubsection{Combination of the \YMy\ and the \MyM\ relations}

We now combine Eq.~\ref{eqn:y_myxmb} with the \MyM\ relation. This
will not change the best-fit parameters, but will increase their
uncertainties. As the determinations of the two relations are
independent, we added quadratically the uncertainties in the best-fit
parameters of the \YMy\ (Eq.~\ref{eq:y_myx}) and
\MyM\ (Eq.~\ref{eq:myx_m}, with values from Eq.~\ref{eq:yx_m})
relations.  Our best-fit \YM\ relation is then
\begin{eqnarray}\label{eq:y_m}
E^{-2/3}(z)\left[\frac{D_{\mathrm{
          A}}^2\,\Yv}{\mathrm{ 10^{-4}\,Mpc^2}}\right] \hspace*{2.5cm} \\
\hspace*{2cm} = 10^{-0.19\pm0.02}
  \left(\frac{(1-b)\,\Mv}{6\times10^{14}\msol}\right)^{1.79\pm0.08}. \nonumber
\end{eqnarray}
Thus inclusion of the statistical uncertainty in the $\Mvy$--$\Mvh$
relation doubles the uncertainty on the normalization and increases
the uncertainty on the slope by $40\%$. {Note that we have implicitly assumed here
that the scatter around the two relations is uncorrelated.}

\subsubsection{Effect of use of an external dataset}

The above results assume a mass estimated from the baseline
$\YX$--$M_{500}$ relation, derived by \citet{arn10} from an external
dataset of 20 relaxed clusters (Eq~\ref{eq:yx_m}). How does this
relation compare to the individual hydrostatic X-ray masses of the
\planck\ cosmological cluster sample?  Of the 71 clusters in the baseline sample:
 
\begin{itemize} 
\item 58 objects have temperature profile information extending to various fractions of $\Rv$, of which
\item 10 cool-core objects have temperature profiles measurements at least out to $\Rv$.
\end{itemize}

Thus, while spatially-resolved
temperature profiles are available for 58 of the 71 clusters with
\xmm\ observations, we must be careful in interpretation of these
data. The \citeauthor{arn10} relation was derived from a carefully
chosen data set consisting of relaxed, cool-core objects having
well-constrained temperature profiles out to around $\Rv$, i.e., the
type of object for which it makes sense to undertake a hydrostatic
mass analysis. Many clusters of the \planck\ sample are merging systems for
which such an analysis would give results that are difficult to
interpret. In addition, few of the \planck\ sample have
spatially-resolved temperature profiles out to $\Rv$. However, as
given in Table~\ref{tab:ym}, the best-fit $\YX$--$\Mvh$ relation for
the 10 cool-core clusters that are detected to $\Rv$ agrees with
Eq.~\ref{eq:y_m} within $1\,\sigma$. Moreover, the relation for the 58
\planck\ clusters with HE mass estimates, derived regardless of dynamical
state  and radial detection extent, also agrees within $1\,\sigma$ (albeit with greatly increased
scatter). We are thus confident that the masses estimated from an
externally-calibrated $\YX$--$\Mvh$ relation are applicable to the
present data set.

\begin{figure*}[]
\begin{center}
\resizebox{0.9\linewidth}{!} {\includegraphics[]{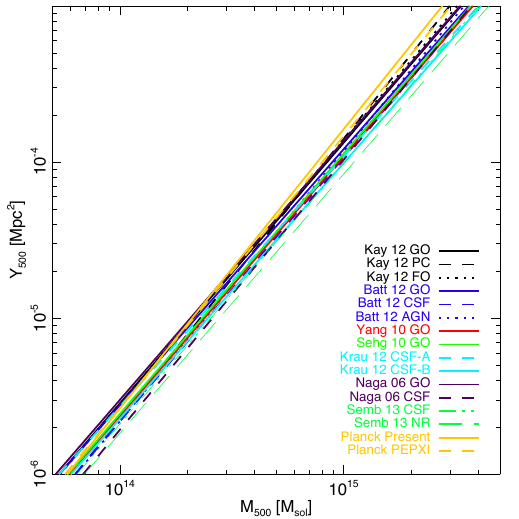}
\hspace{1cm}
\includegraphics[]{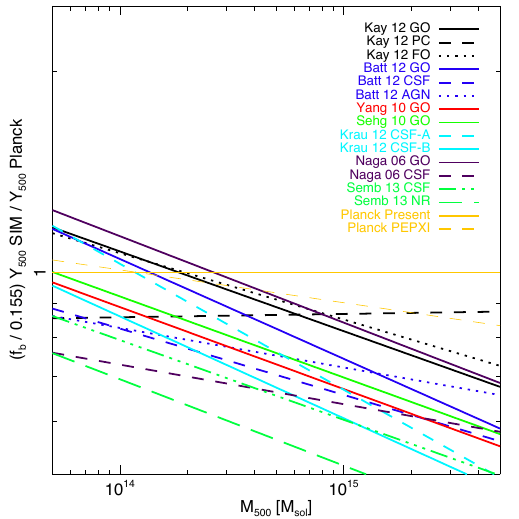}}
\caption{\label{fig:comp_sims} {\it Left:} comparison of
  $\Yv$--$\Mv$ relations from 12 simulations undertaken by six
  different groups with the updated observational
  $\Yv$--$\Mv^{\mathrm{ HE}}$ result from \planck,
  Eq.~\ref{eq:y_m}. {\it Right:} ratio of each simulated
  $\Yv$--$\Mv$ relation  relative to Eq.~\ref{eq:y_m}. The different
  scaling laws are taken from \citet{kay12}, \citet{bat12},
  \citet{yan10}, \citet{seh10}, \citet{kra12}, \citet{nag06}, \citet{sem13} and
  \citet{planck2011-5.2b}.}
\end{center}
\end{figure*}

\subsubsection{Dispersion about the observed relations}

A key issue is the dispersion around the mean relation. 
We first estimate the intrinsic scatter of the \YMh\ relation by combining the intrinsic scatter of the
\YMy\ relation and that of the \MyMh\ relation. This estimate is
applicable to relaxed objects only, since the \YMh\ relation has been
measured using a sample of such systems. If the scatter about the input relations is independent, this gives
\begin{equation}
\sigma  =   \sqrt{\sigma_{\mathrm{ \YSZ\mid\Mvy}}^2 +  2
  \cos^2(\tan^{-1}{\alpha})\,\sigma_{\mathrm{ \Mvh\mid\YX}}^2 }\,,  
\label{eq:scatym}
\end{equation}
where {$\alpha$} is the slope of the \YMy\ relation.  As the HE mass
estimate introduces extra scatter as compared to the true mass
\citep{kay12}, the dispersion about the \YM\ relation is expected to
be smaller than that of the \YMh\ relation {(although this will depend
  on correlations between the scatter in the $\Mvh$--$\Mv$ and
  $\Mvh$--$\Yv$ relations)}. The above expression thus also provides
an estimate of the scatter of the \YM\ relation, again for relaxed
objects.  While merging events are expected to induce shocks in
  the ICM, leading to higher temperatures and thus an increase in
  $\Yv$, current simulations suggest that this is a weak effect. This
  may be due to the relatively short duration of the shocking phase
  during a merger \citep[e.g.,][]{ric01,rit02,poo07}.  Thus, further
assuming that the intrinsic scatter of the \YM\ relation is the same
for the total relaxed and unrelaxed population, as indicated by
numerical simulations \citep{kra06,kay12}, Eq.~\ref{eq:scatym} gives a
conservative estimate of the intrinsic scatter of the \YM\ relation.

The intrinsic dispersion about our baseline $\YX$--$\Mvh$ relation
(Eq.~\ref{eq:yx_m}), taken from \citet{arn10}, is not measurable;
neither is it measurable for the best-fit \chandra\ $\YX$--$\Mvh$
relation published in \citet{vik09b}. Using a smaller sample of 10
systems, \citet{arn07} measured an intrinsic scatter of
$\sigma_{\mathrm{ \log{\Mvh\mid\YX}}}=0.039$ ($9$ percent), in
excellent agreement with the results of the simulations of
\citet{nag07} for the scatter of the \MhYX\ relation for relaxed
clusters ($8.7$ percent, their Table~4). It is somewhat larger than
the intrinsic scatter of the relation between the true mass and $\YX$
derived by Kravtsov et al.\ ($\sigma_{\mathrm{\log{\Mv\mid\YX}}}=5-7$
percent) but close to the results of \citet{fab11}, who find
$\sigma_{\mathrm{ \log{\Mv\mid\YX}}}=$0.036-0.046. We thus take as a
conservative estimate $\sigma_{\mathrm{ \log{\Mvh\mid\YX}}}=0.05$. The
intrinsic dispersion about the $\Yv$--$\Mvy$ relation for our data is
$\sigma_{\mathrm{\log{\YSZ\mid\Mvy}}}=0.065\pm0.01$. This value is
three times larger than the results of \citet{kay12}. Partly this is
due to the presence of outliers in our dataset (as discussed in Paper
1), and it may also be due to projection effects in observed data sets
\citep{kay12}.

Our final observational estimate of the intrinsic scatter is then
$\sigma_{\mathrm{ \log{\YSZ\mid\Mv}}} < 0.074$ or 18 percent, similar
to the predictions from \citet{kay12} and \citet{seh10}.  These
predictions depend both on the numerical scheme and specific physics
assumptions, with values varying by a factor of two in the typical
range $0.04$ to $0.08$ (references in Sect.~\ref{sec:numrec} below).
 
\subsection{Assessing the bias from comparison with numerical simulations}

The final piece of the jigsaw consists of assessing the bias $b$ in
Eq.~\ref{eq:mb}. Since the relation has been calibrated using the HE
mass for a sample of {\it relaxed clusters}, $b$ represents the bias
between $\Mvh$ and the true mass for this category of clusters. In
principle, this can be assessed through comparison with numerical
simulations. However, this approach is hampered by two
difficulties. The first is the exact definition of `relaxed', since it
is almost impossible to select such clusters from observations and
simulations according to the {\it same} criteria. The second is the
specific implementation of the HE equation, which can differ
substantially between observations (e.g., the use of forward fitting using
parametric models, etc.) and simulations (e.g., the use of mock
observations, etc.). Thus the amplitude of the bias that is found will
depend not only on physical departures from HE, but also on technical
details in the approach to data analysis.

Here we use a different approach that avoids these pitfalls, assessing
the bias $b$ by comparing directly the estimated \YM\ relations with
those found from numerical simulations.  We then discuss the
consistency of the resulting bias estimate with the HE bias expected
from simulations and from absolute calibration uncertainties.

\subsubsection{Comparison of simulated $\Yv$--$\Mv$ relations and data}
\label{sec:numrec}

We first compared the $\Yv$--$\Mv$ relations from 14 different
analyses done by seven groups
\citep{nag06,yan10,seh10,kra12,bat12,kay12,sem13}. We translated these
simulations results into a common cosmology and, where necessary,
converted cylindrical relations into spherical measurements assuming a
ratio of $Y_{\mathrm{500, cyl}} /Y_{\mathrm{ 500, sph}} = 0.74/0.61
\simeq 1.2$, as given by the \citet{arn10} pressure profile.

The left-hand panel of Fig.~\ref{fig:comp_sims} shows the different
\mbox{$\Yv$--$\Mv$} relations rescaled to our chosen cosmology. The
simulations use various different types of input physics, and the
resulting $\Yv$--$\Mv$ relations depend strongly on this factor. The
only obvious trend is a mild tendency for adiabatic simulations to
find nearly self-similar slopes (1.66). Runs with non-gravitational
processes tend to find slightly steeper slopes, but this is not always
the case (e.g., the \citealt{kra12} simulations). The right-hand panel
of Fig.~\ref{fig:comp_sims} shows the {\it ratio} of each simulation
$\Yv$--$\Mv$ relation to the \planck\ $\Yv$--$\Mv^{\mathrm{\YX}}$
relation given in Eq.~\ref{eq:y_m}. All results have been rescaled to
account for the differences in baryon fraction between simulations. At
our reference pivot point of $M_{500} = 6 \times 10^{14}\, \msol$, all
simulations are offset from the measured relation. There is also a
clear dependence on mass arising from the difference in slope between
the majority of the simulated relations and that of the
\planck\ relation. The \planck\ slope is steeper, possibly indicating
the stronger effect of non-gravitational processes in the real data.

\subsubsection{Quantification of the mass bias}

We define the mass bias $b$ between the `true' and observed $\Mv$
values, following Eq.~\ref{eq:mb} and explicitly allowing for possible
mass dependence of the bias, i.e, $b=b(\Mv^{\mathrm{true}})$.  Both
masses are defined at a fixed density contrast of $500$, so that the
relations between observed and true mass and radius read
\begin{eqnarray}\label{eq:mot}
%\Mv^{\mathrm{  obs}} & =&  (1-b)\,\Mv^{\mathrm{ true}}\,, \\ \nonumber
%\Rv^{\mathrm{ obs}} & =&  (1-b)^{1/3}\,\Rv^{\mathrm{ true}} \,,
\Mv^{\mathrm{  obs}} & =&  \left[1-b\, \left(\Mv^{\mathrm{true}} \right) \right] \,\Mv^{\mathrm{ true}}\, \\  
 \Rv^{\mathrm{ obs}} & =&  \left[1-b\, \left(\Mv^{\mathrm{true}} \right) \right]^{1/3}\,\Rv^{\mathrm{ true}} \,  %  \nonumber
\end{eqnarray}
where `true' denotes simulated quantities, and `obs' denotes
quantities estimated at the apertures derived from observations.  
 The corresponding \YM\ relations are
 \begin{eqnarray}\label{eq:ymt}
Y \left(<\Rv^{\mathrm{true}} \right) & = & A_{\mathrm{true}} \left[\Mv^{\mathrm{true}}\right]^{\beta}\,, \\ %  \nonumber
\label{eq:ymo}Y \left(<\Rv^{\mathrm{obs}}\right) & = & A_{\mathrm{obs}} \left[\Mv^{\mathrm{obs}} \right]^{\alpha}\, %, \\ \nonumber
 \end{eqnarray}

In our case, $Y_{500}$ is measured interior to $\Rvy$ as opposed to
$\Rv^{\mathrm{true}}$. 
The ratio $Y \left(<\Rv^{\mathrm{true}}\right)/ Y \left(<\Rv^{\mathrm{obs}}\right) $ depends on the radial variation of $\YSZ$ for scaled
radii, $r/\Rvy=\Rv^{\mathrm{ true}} /\Rv^{\mathrm{ obs}}=
(1-b)^{-1/3}$, which is close to $1$.  For a GNFW universal profile
\citep{arn10}, we find that it  can be well fit by a power law of
the form $(1-b)^{-1/4}$.  Combining Eq.~\ref{eq:mot},  Eq.~\ref{eq:ymt} and   Eq.~\ref{eq:ymo} we then arrive at
\begin{equation}\label{eq:by}
\left[1-b\, \left(\Mv^{\mathrm{true}}\right)\right]=\left[ \frac{   A_{\mathrm{true}}   \left(\Mv^{\mathrm{true}}  \right)^{\beta}   } { A_{\mathrm{obs}}   \left(\Mv^{\mathrm{true}}  \right)^{\alpha}    }  \right]^{-1/4+\alpha}\,.
\end{equation}
The relation makes it clear that a mass dependence of the bias
naturally translates into a different slope of the observed and true
\YM\ relations.

\begin{figure}[]
\begin{center}
\includegraphics[width=0.85\columnwidth]{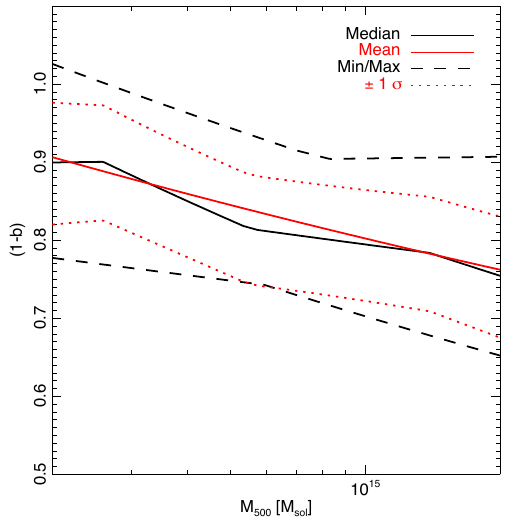}
\caption{\label{fig:bvM} The dependence of $(1-b)$ on mass. Note that
  this value is strongly dependent on the baryon fraction
  $f_\mathrm{b}$ (see text).}
\end{center}
\end{figure}

The bias $b$ can then be estimated from a comparison of observed and
simulated relations, with the caveat that differences can also arise
from imperfect modelling of cluster physics within the simulations. For
the ensemble of simulations shown in Fig.~\ref{fig:comp_sims}, the
right panel shows the ratio of the observed and simulated relations as
a function of mass.  Figure~\ref{fig:bvM} shows the corresponding
variation of $(1-b)$ as a function of mass $M$ from Eq.~\ref{eq:by}
for the observed slope $\alpha=1.79$.  This is mass dependent due to
the difference in slopes between the simulated and observed
relations. At a pivot point of $\Mv = 6 \times 10^{14}\, \msol$, the
median value of $A_{\mathrm{ true}} / A_{\mathrm{ obs}}$ is $0.74$,
implying $(1-b) = 0.81$. However, there is a large amount of scatter
in the predictions from simulations. As a consequence, $(1-b)$ can
vary from 0.74 to 0.97 at $\Mv = 6 \times 10^{14}\, \msol$. Note that
the above results depend significantly on the baryon fraction
$f_\mathrm{b}$. For example, assuming the WMAP-7 value
$f_\mathrm{b} = 0.167$, the median value of $(1-b)$ is $0.86$ at the
pivot point of $\Mv = 6 \times 10^{14}\, \msol$.

\subsubsection{Consistency with HE bias predictions and absolute
  calibration uncertainties} 

Taken at face value, the bias we derive above of $(1-b)\simeq 0.8$
implies that the HE mass used to calibrate the \YMh\ relation is
offset from the true mass by around $20$ percent. Is this reasonable?

We can first compare HE X-ray and weak lensing (WL) masses.  Although
as mentioned above both measurements are expected to be biased, such
comparisons are useful because the mass measurements involved are
essentially independent.  In addition measurements for moderately
large sample sizes (tens of systems) are now starting to appear in the
literature.  However, at present there is little consensus, with some
studies finding good agreement between HE and WL masses
\citep[e.g.,][]{vik09b,zha10}, some finding that HE masses are lower
than WL masses, \citep[e.g.,][]{mah08}, and some even finding that HE
masses are higher than WL masses \citep[][]{planck2012-III}. The key
point in such analyses is rigorous data quality on both the X-ray and
optical sides. Most recent work points to relatively good
agreement between X-ray and WL masses, with $M^{\mathrm{ HE}}/
M^{\mathrm{ WL}} \simeq 0.9$ on average, and $M^{\mathrm{
    HE}}/M^{\mathrm{ WL}} \simeq 1$ for relaxed systems
\citep{mah12,von12}.

According to cosmological numerical simulations, the measurement bias
induced by X-ray measurements relative to the `true' values can be
caused by two main effects. The first is the classic `hydrostatic mass
bias' due to non-thermal pressure support from turbulence/random
motions, etc. However, the exact details are very model-dependent. The
HE bias expected from simulations varies substantially, depending on
the details of the numerical scheme, the input physics, and the
approach used to calculate the HE masses \citep[e.g.,][]{ras12}. In
addition, the amount of bias is different depending on the dynamical
state of the object, relaxed systems having less bias than unrelaxed
systems. The majority of numerical simulations predict HE biases of 10
to 20 percent \citep{nag07,pif08,lau09,kay12,ras12}.

Temperature inhomogeneities constitute the second contributor to X-ray
measurement bias. In the presence of large amounts of cool gas, a
single-temperature fit to a multi-temperature plasma will yield a
result that is biased towards lower temperatures
\citep[e.g.,][]{maz04}. The presence of temperature inhomogeneities
will depend on the dynamical state. While this effect can be
investigated with simulations, estimates of its impact vary widely,
owing to differences in numerical schemes and the different
implementations of the input physics. For instance, simulations with
heat conduction consistently predict smoother temperature
distributions, thus X-ray spectroscopic biases are minimal in this
case. On the other hand, `adiabatic' simulations predict long-lasting
high-density cool-core type phenomena, which will lead to significant
biases in single-temperature fits. Estimates of biasing due to
temperature inhomogeneities can range up to 10 or 15 percent
\citep[e.g.,][]{ras12}.

Finally, for HE mass estimates obtained from X-ray data, instrument
calibration uncertainties also play a significant role in introducing
uncertainties in mass estimates. For instance, the difference in
calibration between \xmm\ and \chandra\ can induce differences in
$\YX$. This is typically $5$ percent, from a comparison of \xmm\ based
values published by \citet{planck2011-5.2b} to \chandra\ values for 28
ESZ clusters by \citet{roz12}. This can lead to differences of up to
10 percent in the mass $\Mvy$ derived from $\YX$, owing to the
dependence of the mass on $\YX$.

Thus our adopted baseline value of $(1-b)\simeq 0.8$, ranging from
$0.7$--$1$, appears to encompass our current ignorance of the exact
bias.

\subsection{Conclusions}

In summary the baseline is 
\begin{equation}
E^{-2/3}(z)\left[\frac{D_{\mathrm{ A}}^2\,\YSZ}
  {\mathrm{10^{-4}\,Mpc^2}}\right] =
10^{-0.19\pm0.02}\left[\frac{(1-b)\Mv}{6\times10^{14}\msol}
\right]^{1.79\pm0.08}\,,
\end{equation}
with an intrinsic scatter of $\sigma_{\log Y}=0.075$ and a mean bias
$(1-b)=0.80_{-0.1}^{+0.2}$. The statistical uncertainty on the
normalization is about $5\%$ and the error budget is dominated
by the systematic uncertainties.

\raggedright 
\end{document}